\documentclass{aa}

\usepackage{txfonts}
\usepackage{graphicx}
\usepackage{natbib}
\usepackage[english]{babel}
\usepackage{aalongtable}
\usepackage[figuresright]{rotating}

\begin{document}
\title{XMM-Newton survey of two Upper Scorpius regions}
\author{C.~Argiroffi\inst{1} \and F.~Favata\inst{2} \and E.~Flaccomio\inst{3} \and A.~Maggio\inst{3} \and G.~Micela\inst{3} \and G.~Peres\inst{1} \and S.~Sciortino\inst{3}}
\offprints{C.~Argiroffi, {\email argi@astropa.unipa.it}}
\institute{Dipartimento di Scienze Fisiche ed Astronomiche, Sezione di Astronomia, Universit\`a di Palermo, Piazza del Parlamento 1, 90134 Palermo, Italy, \email{argi@astropa.unipa.it, peres@astropa.unipa.it} \and Astrophysics Division - Research and Science Support Department of ESA, Postbus 299, 2200 AG Noordwijk, The Netherlands \email{Fabio.Favata@rssd.esa.int} \and INAF - Osservatorio Astronomico di Palermo, Piazza del Parlamento 1, 90134 Palermo, Italy, \email{ettoref@astropa.unipa.it, maggio@astropa.unipa.it, giusi@astropa.unipa.it, sciorti@astropa.unipa.it}}
\date{Received 23 May 2006 / Accepted 11 July 2006}
\titlerunning{XMM-Newton observations of Upper Scorpius}
\authorrunning{C.~Argiroffi et al.}
\abstract
{}
{We study X-ray emission from young stars by analyzing deep XMM-Newton observations of two regions of the Upper Scorpius association, having an age of 5\,Myr.}
{Based on near infrared and optical photometry we identify 22 Upper Scorpius photometric members among the 224 detected X-ray sources. We derive coronal properties of Upper Scorpius stars by performing X-ray spectral and timing analysis. The study of four strong and isolated stellar flares allows us to derive the length of the flaring loops.}
{Among the 22 Upper Scorpius stars, 13 are identified as Upper Scorpius photometric members for the first time. The sample includes 7 weak-line T~Tauri stars and 1 classical T~Tauri star, while the nature of the remaining sources is unknown. Except for the intermediate mass star HD~142578, all the detected USco sources are low mass stars of spectral type ranging from G to late M. The X-ray emission spectrum of the most intense Upper Scorpius sources indicates metal depleted plasma with temperature of $\sim10$\,MK, resembling the typical coronal emission of active main sequence stars. At least 59\% of the detected members of the association have variable X-ray emission, and the flaring coronal structures appear shorter than or comparable to the stellar radii already at the Upper Scorpius age. We also find indication of increasing plasma metallicity (up to a factor 20) during strong flares. We identify a new galaxy cluster among the 224 X-ray source detected: the X-ray spectrum of its intra cluster medium indicates a redshift of $\sim0.41\pm0.02$.}
{}
\keywords{stars: abundances -- stars: activity -- stars: coronae -- stars: flare -- stars: pre-main sequence -- X-rays: stars}
\maketitle

\section{Introduction}

Pre-main sequence (PMS) low-mass stars at an age of a few Myr are usually classified as classical T~Tauri stars (CTTSs) if they still accrete material from their circumstellar disk; they become weak-line T~Tauri stars (WTTSs) when the accretion ends. This classification is usually based on the H$\alpha$ line \citep[e.g.][and references therein]{HartmannHewett1994,WhiteBasri2003}.

Strong X-ray radiation is a common characteristic of PMS low-mass stars \citep[e.g.][and references therein]{FeigelsonMontmerle1999}. X-ray emission from accreting and non-accreting PMS stars is produced by hot plasma ($T\sim1-100$\,MK) suggesting therefore that coronal activity is present irrespective of the accretion status. However low resolution X-ray spectra hinted at significant differences between CTTSs and WTTSs. CTTSs reveal lower X-ray luminosity, with respect to the non-accreting WTTSs, moreover CTTSs appear to produce X-ray spectra harder than WTTSs \citep{NeuhaeuserSterzik1995,TsujimotoKoyama2002,FlaccomioMicela2003a,PreibischKim2005,FlaccomioMicela2006}. These results indicate that accretion processes and circumstellar disks do affect the X-ray emission of PMS stars. How this influence occurs is still a debated question. During the CTTS stage the infalling circumstellar material, heated by the shock with the photosphere to temperatures of few MK, is a possible source of soft X-ray emission \citep{KastnerHuenemoerder2002}. Also protostellar jets  may contribute to the X-ray emission of PMS stars \citep{PravdoFeigelson2001,FavataFridlund2002}. Long coronal loops found in PMS stars \citep{FavataFlaccomio2005} suggest that interactions between the magnetosphere and the circumstellar disk may occur and affect the coronal plasma characteristics. On the other hand the lack of accretion streams and close circumstellar material in WTTSs imply that their X-ray radiation may only be produced by magnetically confined coronal plasma.

The different rotational periods of CTTSs and WTTSs, evidenced by \citet{BouvierCabrit1993}, may explain the differences in the X-ray properties, as it happens for main sequence stars \citep{PallaviciniGolub1981,PizzolatoMaggio2003}. However several studies suggested that the X-ray luminosity vs. rotational period relationship is not so clear for PMS stars as in the case of MS stars, possibly because most of the PMS stars appear to be in the saturated or supersaturated regime, or because other parameters, like stellar age, may alter the X-ray activity level \citep{StelzerNeuhaeuser2001,FlaccomioMicela2003b,StassunArdila2004,PreibischFeigelson2005,PreibischKim2005}.

The analysis of X-ray emission from WTTSs, where the lack of close circumstellar material limits corona-disk interactions and excludes accretion driven X-ray emission, can provide insightful results on the comprehension of coronal emission from young stars. These findings may also be useful to understand the properties of coronal plasma of CTTSs. To this aim we have analyzed two {\it XMM-Newton} observations of two fields in the Upper Scorpius association, the younger portion of the Scorpius Centaurus OB association.

\begin{table*}
\begin{center}
\caption{Log of the two {\it XMM} observations of Upper Scorpius star forming region. Filtered exposures refer to screened time intervals, selected because not affected by high background count rates.}
\label{tab:log}
\begin{tabular}{lccccc}
\hline\hline
EPIC       & Science      & Optical Blocking & \multicolumn{2}{c}{Exposure (ks)} & Observation Start    \\
Instrument & Mode         & Filter           & Total    & Filtered               & (UT)                 \\
\hline
\multicolumn{6}{c}{Upper Scorpius - field 1 - rev. 130 - ObsId 0109060201 - RA=16:14:00.0 DEC=$-$23:00:00.0} \\
\hline
PN         & Full Frame   & Medium           & 53.3     & 45.7                   & 2000 Aug 24 21:04:08 \\
MOS1       & Full Frame   & Medium           & 53.3     & 50.0                   & 2000 Aug 24 20:23:08 \\
MOS2       & Full Frame   & Medium           & 52.8     & 50.4                   & 2000 Aug 24 20:23:05 \\
\hline
\multicolumn{6}{c}{Upper Scorpius - field 2 - rev. 131 - ObsId 0112380101 - RA=15:56:25.0 DEC=$-$23:37:47.0} \\
\hline
PN         & Full Frame   & Medium           & 42.7     & 36.3                   & 2000 Aug 26 23:41:37 \\
MOS1       & Full Frame   & Medium           & 43.8     & 39.8                   & 2000 Aug 26 23:00:43 \\
MOS2       & Full Frame   & Medium           & 43.8     & 40.0                   & 2000 Aug 26 23:00:37 \\
\hline
\end{tabular}
\end{center}
\end{table*}

The Scorpius Centaurus OB association is one of the most nearby regions of recent star formation. It contains an estimated population of $\sim 5\,000-10\,000$ stars with mass larger than $0.1\,M_{\sun}$ \citep{deGeus1992}. It is the result of a star formation process started $\sim 15$\,Myr ago in a giant molecular cloud. The Scorpius Centaurus OB association is composed of three subgroups: the Upper Scorpius (USco), the Upper Centaurus Lupus (UCL), and the Lower Centaurus Crux (LCC). These three subgroups have different ages (ranging from 5 to 15\,Myr) suggesting that star formation proceeded progressively in different parts of the original molecular cloud. According to the proposed scenario the older OB stars, located in the UCL subgroup, produced shock fronts in the original giant molecular cloud by their strong winds and subsequent supernova explosions, and hence triggered new star formations \citep{deGeus1992}. These shock waves passed $\sim5$\,Myr ago in the USco region activating the USco star formation. Once the most massive USco stars evolved, $\sim1.5$\,Myr ago, their supernova explosions dissolved the interstellar material in the USco region hence stopping the star formation process. Therefore the USco region appears free of dense interstellar material. Moreover the shocks produced by massive USco stars induced the latest stage of the star formation process in the $\rho$~Ophiuchi cloud which lies at the margin of the Scorpius Centaurus OB association.

\begin{table*}
\begin{center}
\caption{Photometric observations of Upper Scorpius selected fields. We report in columns: the field center, the size of the FOV, the instrument, and the broad band filters employed.}
\label{tab:phot}
\begin{tabular}[h]{ccccc}
\hline\hline
RA (J2000) & DEC (J2000) & FOV area                                     & Instrument     & Bands           \\
\hline
15:56:20.8 & -23:31:60 & $1\degr9\arcmin\times1\degr9\arcmin$         & Curtis Schmidt & $U B V R_c I_c$ \\
15:58:52.0 & -22:49:47 &  "                                           &  "             & "               \\        
15:56:21.5 & -23:37:27 & $13\arcmin40\arcsec\times13\arcmin40\arcsec$ & Danish 1.54m   & $V R_c I_c$     \\      
15:56:21.4 & -23:52:28 &  "                                           &  "             & "               \\
15:55:00.5 & -24:00:52 &  "                                           &  "             & "               \\
15:55:00.5 & -23:47:41 &  "                                           &  "             & "               \\
15:55:19.4 & -23:17:04 &  "                                           &  "             & "               \\
15:54:09.4 & -23:21:16 &  "                                           &  "             & "               \\
15:58:05.3 & -23:06:53 &  "                                           &  "             & "               \\
\hline
\end{tabular}
\end{center}
\end{table*}

Among the three subgroups of Scorpius Centaurus OB association the USco is the youngest. Its age, estimated from high, intermediate, and low-mass stars, is $\sim5$\,Myr \citep{deGeusdeZeeuw1989,PreibischBrown2002,SlesnickCarpenter2006}. Measured parallaxes of USco members suggest an average distance of $\sim145$\,pc \citep{deBruijneHoogerwerf1997}, with a scatter of $\pm20$\,pc. The USco region displays a CTTSs vs. WTTSs ratio of $\sim2-5\%$ \citep{WalterVrba1994,Martin1998}, which is lower than the ratio observed in T~associations of similar age. The supernova shock fronts, which propagated in the USco region $\sim1.5$\,Myr ago, might have cleaned the local medium, and accelerated the dissipation of circumstellar disks of accreting stars \citep{WalterVrba1994}. The vicinity of the USco region and its low circumstellar extinction \citep[$A_{V}\la2$\,mag,][]{deGeusdeZeeuw1989} offer the opportunity to perform detailed studies of the X-ray emission of T~Tauri stars, and in particular of WTTSs, abundant in this region.

In this paper we present the analysis of {\it XMM-Newton} observations of two fields of the USco region. In order to select probable PMS candidates of USco we have used near infrared (NIR) data of 2MASS and DENIS catalogs. We stress that the 2MASS catalog, thanks to its sky coverage and limiting magnitude, permits complete detection of USco sources down to $\sim0.02\,M_{\sun}$. Moreover we have also taken advantage of optical photometry \citep{FlaccomioFavata2000} obtained with two instruments: the Curtis Schmidt (CTIO) telescope, and the Danish 1.54m (ESO) telescope with the Danish Faint Object Spectrograph and Camera.


\section{Observations and data analysis}

\subsection{X-ray observations}
\label{xobs}

{\it XMM-Newton} observed two regions of the USco association on 2000 August 24 and 26, with exposures of $\sim53$ and $\sim43$\,ks. For both the observations and for all the EPIC instruments the medium filters were used to prevent optical contamination in the softer part of the spectra. The log of the {\it XMM-Newton} observations is reported in Table~\ref{tab:log}.

We have processed both observations using the SAS.V6.0 standard tasks {\tt epchain} and {\tt emchain}. We have selected events with energy ranging from 0.3 to 7.9\,keV: below 0.3\,keV the instruments response is not well calibrated, above 7.9\,keV background dominates over the stellar X-ray emission.

To maximize the $S/N$ we have performed time screening of the X-ray data discarding time intervals affected by high background count rates. The screened good time intervals correspond to 85-95\% of the initial exposures, while the background count rate of the EPIC/PN instrument is reduced by a factor $\sim2$.

\subsection{Optical photometric observations}

The photometric observations were performed with the CTIO Curtis-Schmidt and Danish 1.54m telescopes. The observed fields of view for each instrument and the relevant bands are reported in Table~\ref{tab:phot}. Note that the fields of view do not cover entirely the two {\it XMM-Newton} observations (see Sect.~\ref{ident}). Data reduction has been carried out using IRAF tools. We have performed aperture photometry on all fields and transformed the instrumental magnitude to the standard Johnson-Cousin system through the use of the \citet{Landolt1992} standard catalog. 

For each field we have either two or three distinct observations performed with the same filter. In order to estimate the uncertainties of our photometry as a function of star brightness we have calculated, for each star, the dispersion of the magnitudes independently measured and have taken the average of such dispersions. These uncertainties are consistent with those formally derived combining the statistical errors on the instrumental magnitudes and the uncertainties on the transformations between instrumental magnitudes to the standard system.

We have considered only objects with (mean) uncertainties on $V$, $I$ and $V-I$ less than $\sim 0.1$ mag. This translates into the limiting magnitudes: $V \le 19.0$ for the Danish 1.54m, $V \le 16.5$ for the Curtis Schmidt. Source detected and their photometry are listed in Tables~B.1 and C.1.

\subsection{X-ray source detection}

\begin{figure*}
\centering
\includegraphics[width=17cm]{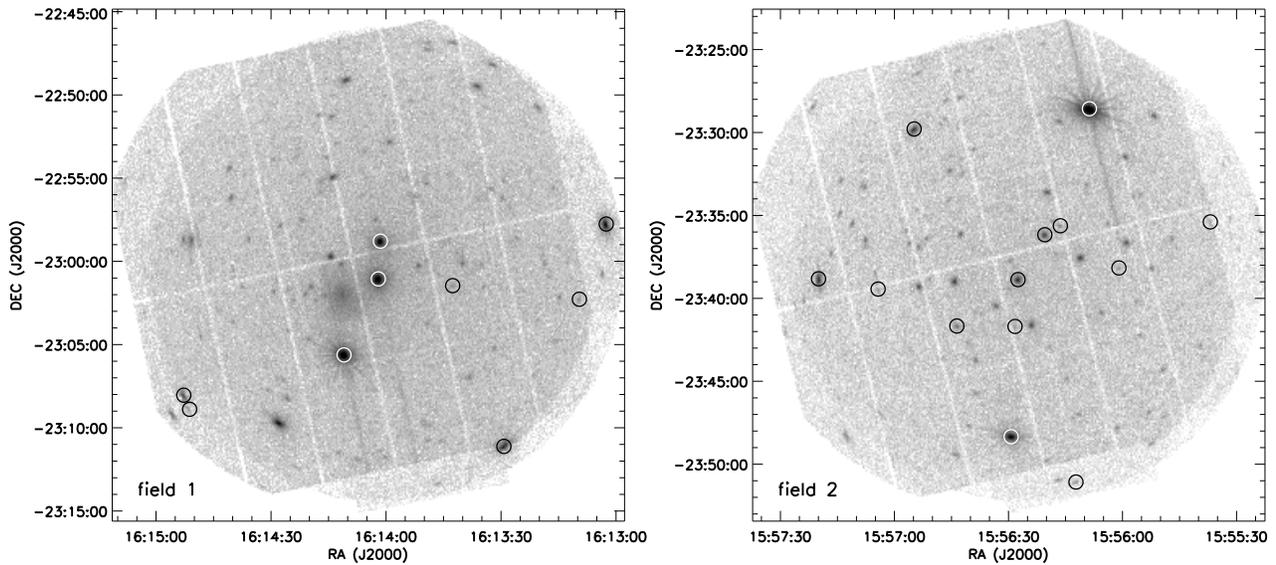}
\caption{Composite images (PN+MOS1+MOS2) of the two {\it XMM-Newton} observations. Circles mark X-ray sources selected as photometric USco members.}
\label{fig:XMMfov}
\end{figure*}

We have searched for X-ray sources in the two observations applying the Wavelet Transform detection algorithm developed for {\it ROSAT} data \citep{DamianiMaggio1997a,DamianiMaggio1997b} and adapted to the EPIC case \citep{PillitteriMicela2006}. We have performed X-ray source detection on composite count rate images obtained by the superposition of the three EPIC instruments, shown in Fig.~\ref{fig:XMMfov}. We have established significance thresholds of 4.8 and 4.9 $\sigma$, for fields 1 and 2 respectively, in order to obtain on average one spurious source per field due to background fluctuations\footnote{The threshold levels were determined applying the following procedure: we have simulated empty fields characterized by the same background levels of the two USco {\it XMM-Newton} observations; we have performed source detections with different threshold levels; we have estimated the threshold which provides the requested number of spurious sources. The derived threshold depends on the background levels, therefore we adopted different threshold levels for the two observations.}. Spurious detections due to peculiar point spread function features or to out of time event signatures are however present but are easily recognized by inspection and discarded. We have also removed from the list spurious sources due to hot pixels. We have obtained 117 and 107 sources in the fields 1 and 2 respectively. The list of the detected {\it XMM-Newton} sources is reported in Table~\ref{tab:src}.

\subsection{Extraction of X-ray data}

For each detected source we have extracted events from a circle centered on the source position. The extraction radii vary from source to source in order to maximize the $S/N$ ratio and to avoid inclusion of other nearby sources. They range from $10\arcsec$, for sources very close to each other, to $60\arcsec$ for intense and isolated sources. Background events have been extracted from annular regions centered on the source or from a circular region located near the source (when possible at the same off-axis and on the same CCD); usually the background region has been chosen to be larger than the source extraction region. For each source an ancillary response function and a redistribution matrix function have been computed using SAS tasks. Source spectra have been rebinned to obtain at least 20 counts per bin. We have verified that EPIC data were in all cases not affected by significant pile-up. Analysis have been performed considering single, double, triple or quadruple pixel events ({\sc pattern} $\le 12$), except for PN spectral analysis where only single or double pixel events ({\sc pattern} $\le 4$) have been retained.

\begin{figure*}
\centering
\includegraphics[width=17cm]{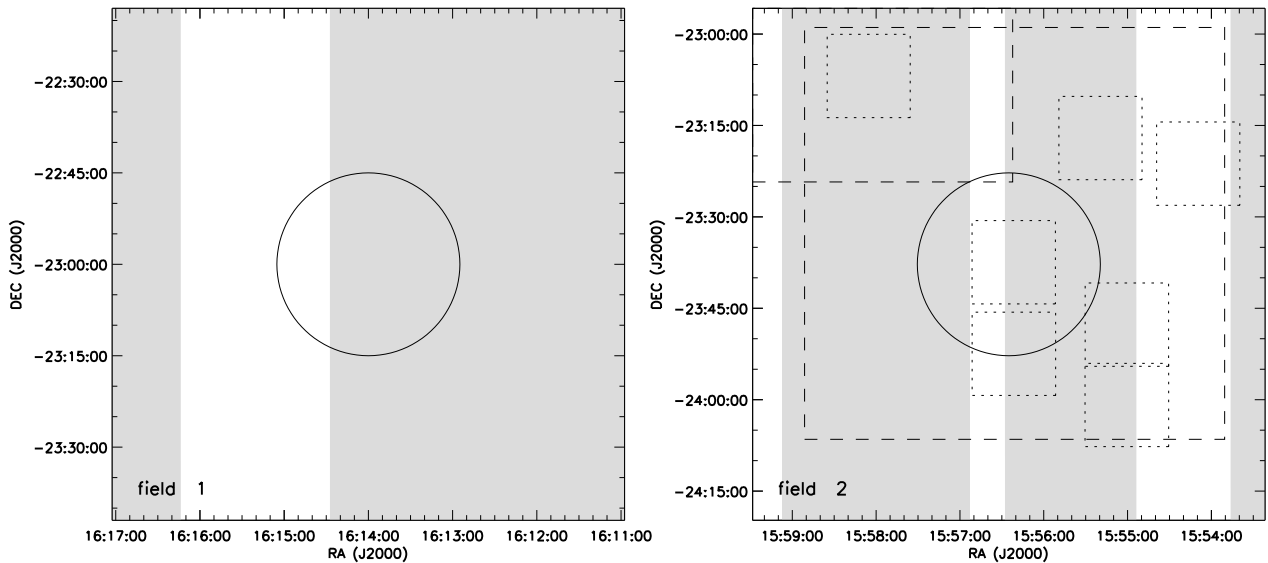}
\caption{Schematic maps of the different photometric surveys coverage of the two observed USco fields. Circles indicate the {\it XMM-Newton} field of view; gray color identifies the area covered by the DENIS catalog; dashed square marks the CTIO observation; short-dashed squares mark the Danish 1.54m fields.}
\label{fig:CATfov}
\end{figure*}


\begin{figure*}
\resizebox{0.5\hsize}{!}{\includegraphics{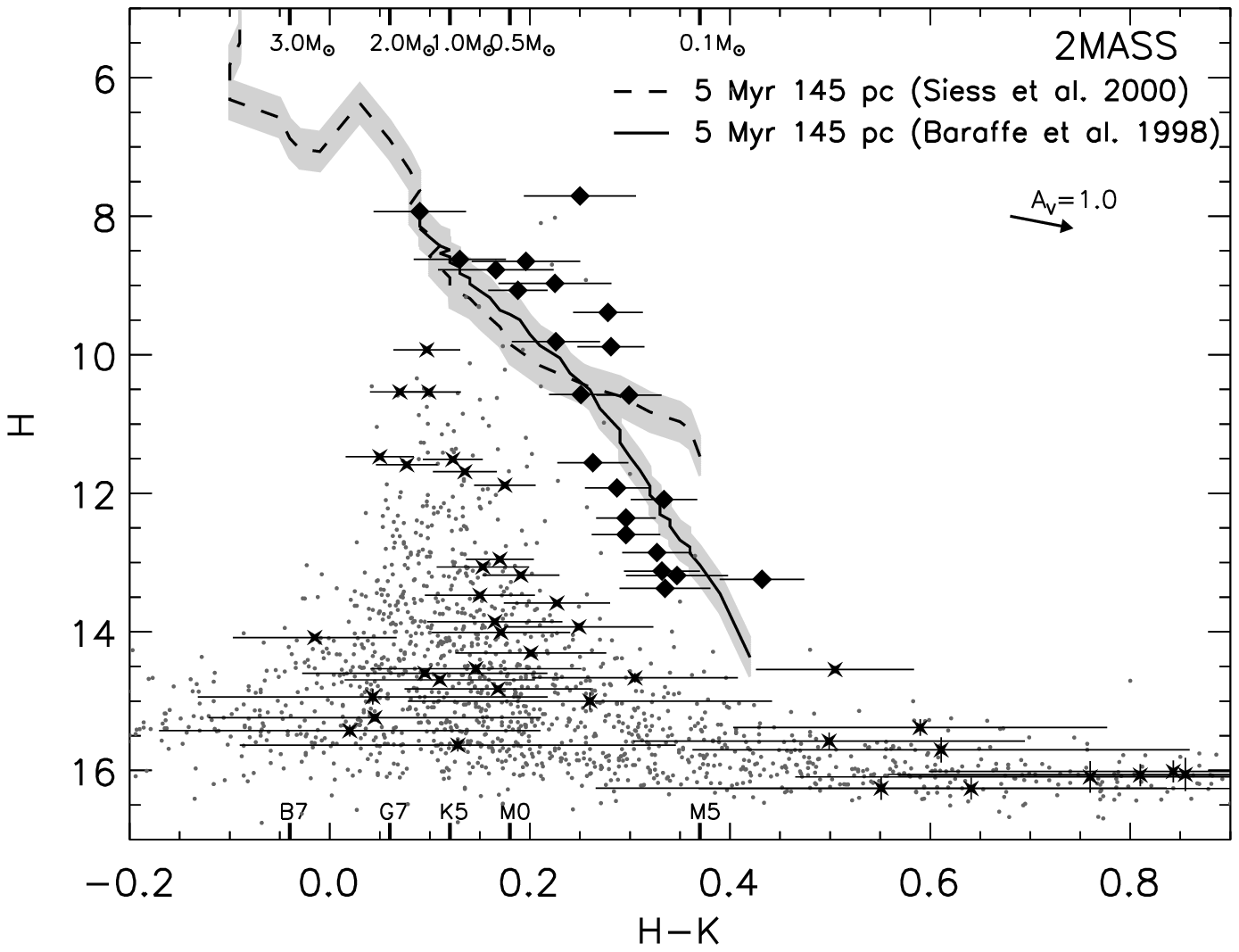}}
\resizebox{0.5\hsize}{!}{\includegraphics{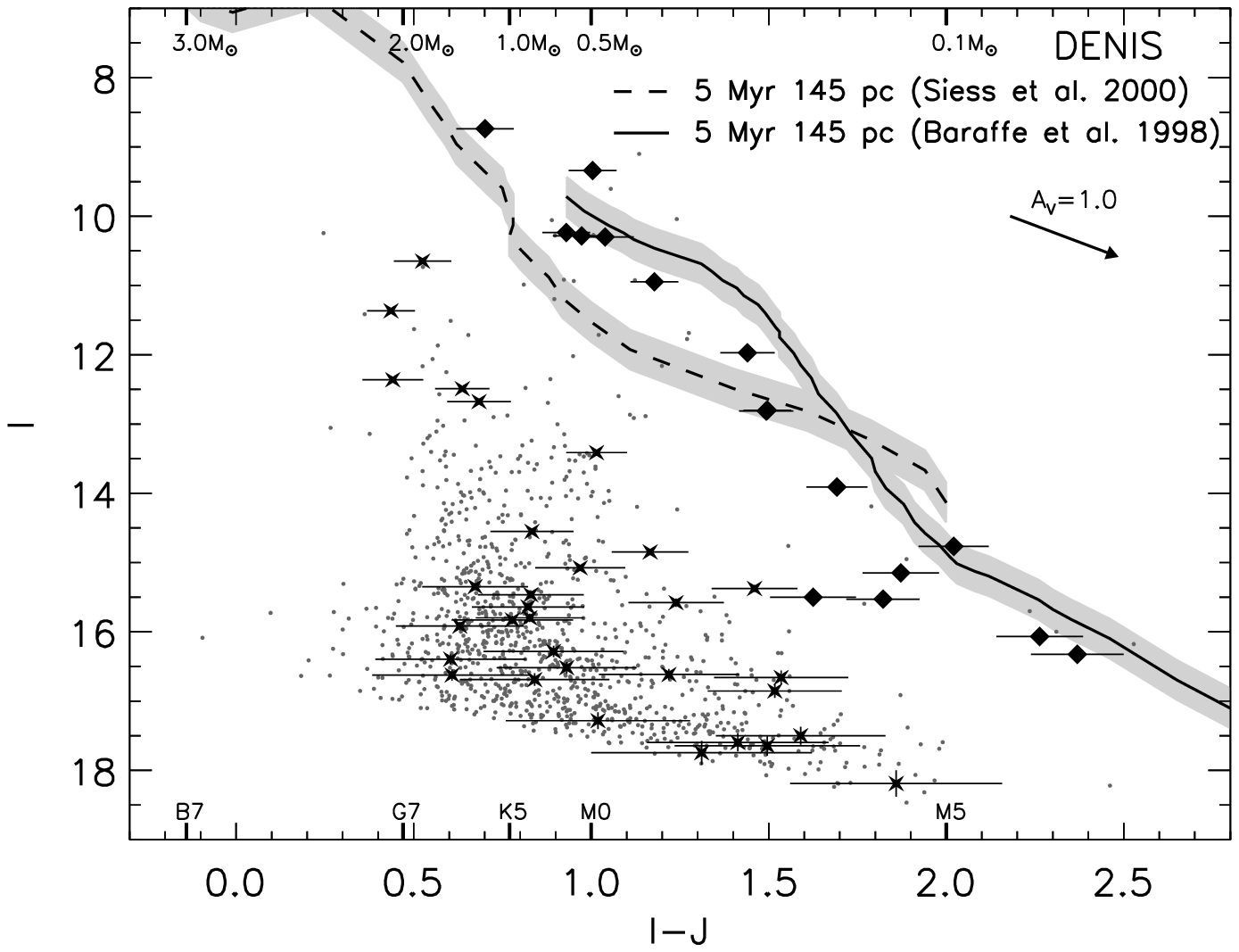}}

\resizebox{0.5\hsize}{!}{\includegraphics{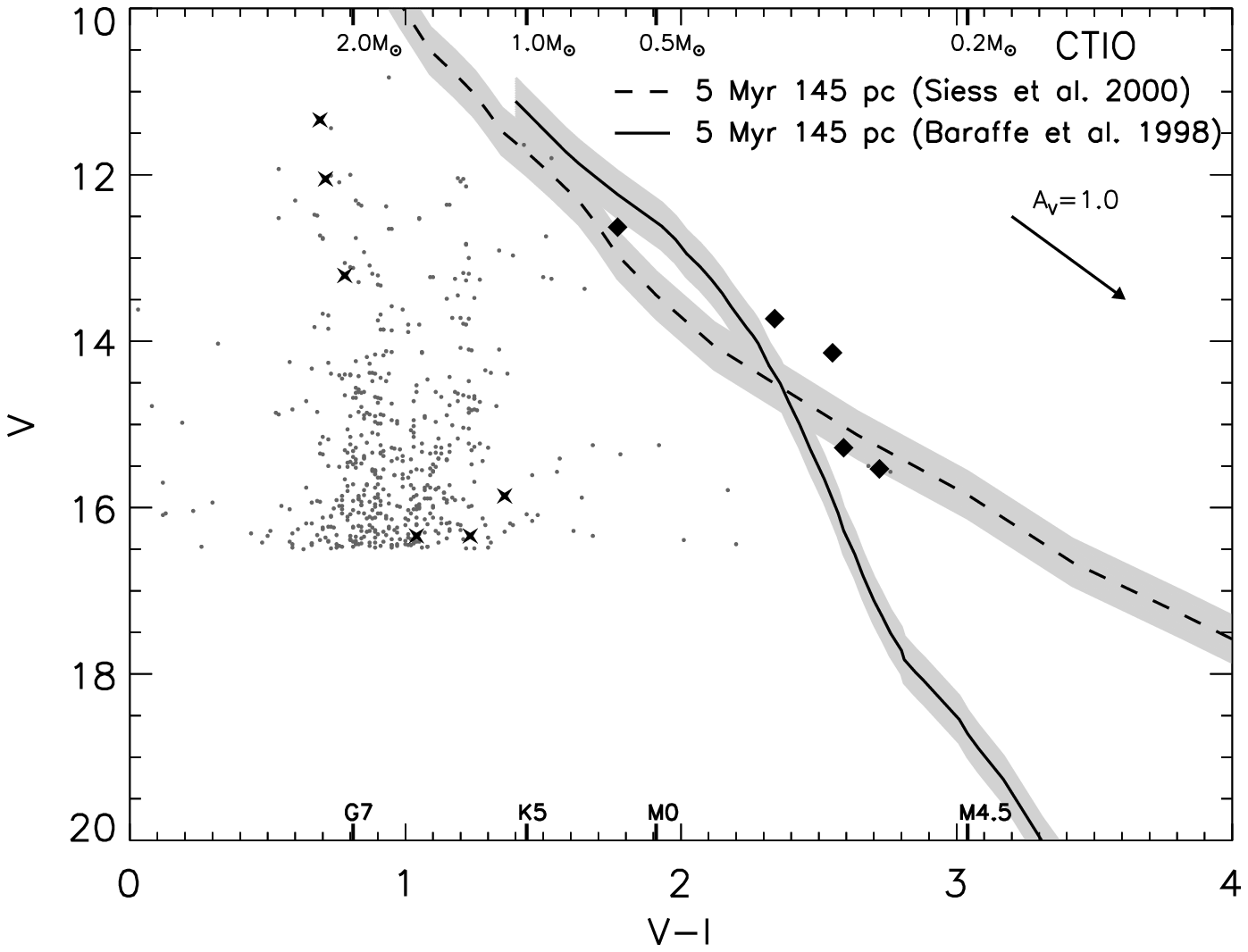}}
\resizebox{0.5\hsize}{!}{\includegraphics{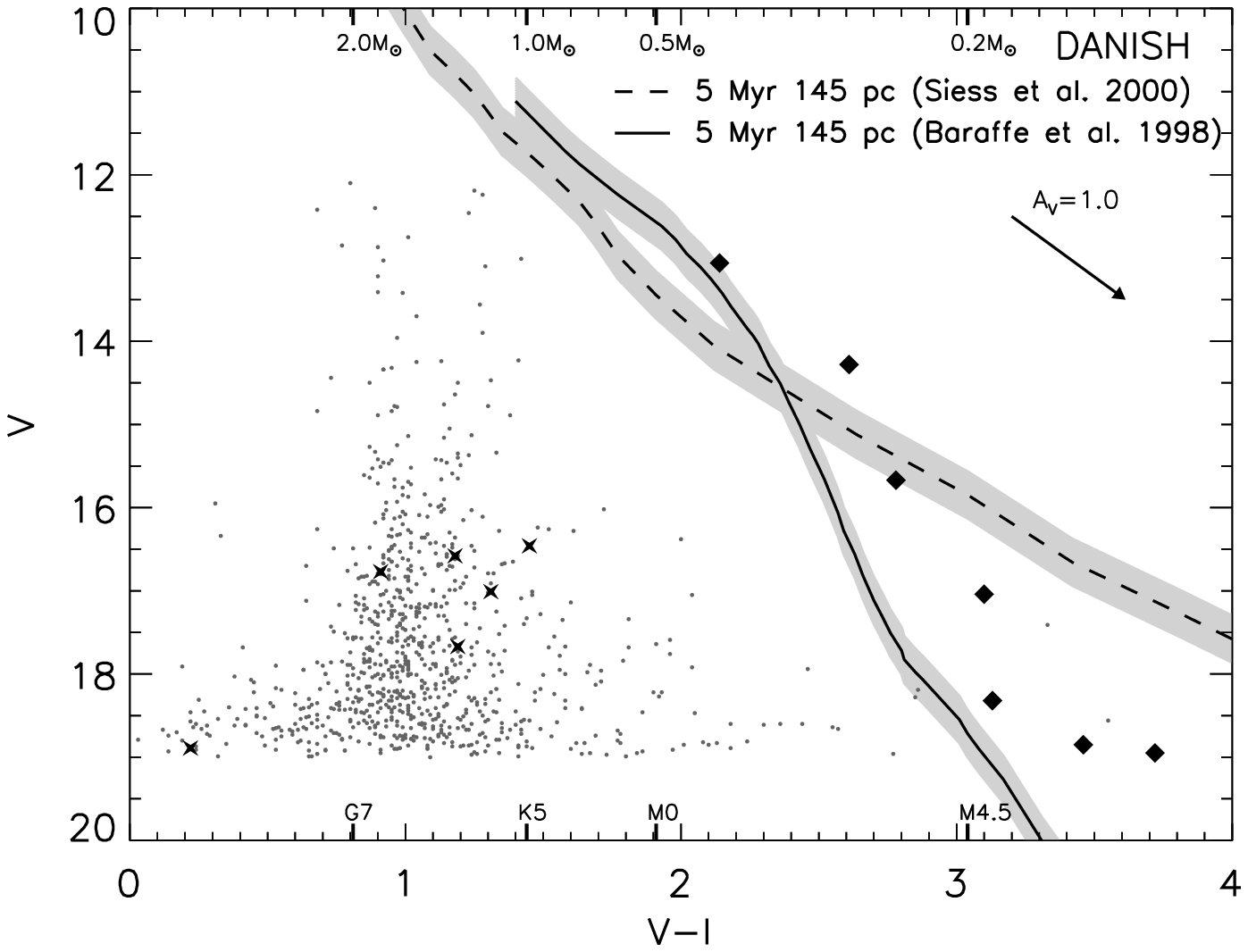}}
\caption{Color magnitude diagrams of the 2MASS, DENIS, CTIO, and Danish 1.54m sources which fall in the two {\it XMM-Newton} fields. In all the plots filled diamonds indicate optical/NIR sources, with X-ray counterparts, selected as photometric USco members. Crosses mark optical/NIR sources, with X-ray counterparts, which are probable background objects. The two isochrones, plotted with dashed and solid lines, are from \protect{\citet{SiessDufour2000}} and \protect{\citet{BaraffeChabrier1998}}, respectively. They correspond to an age of 5\,Myr and are scaled to the distance of 145\,pc. The \protect{\citeauthor{SiessDufour2000}} model ranges from 0.1 to $7.0\,M_{\sun}$, while the \protect{\citeauthor{BaraffeChabrier1998}} spans the $0.02-1.4\,M_{\sun}$ interval. The light gray area identify the locus obtained by varying the distance of $\pm 20$\,pc. Masses and spectral types reported in the upper and lower part of the plot refer to the \protect{\citeauthor{SiessDufour2000}} isochrone.}
\label{fig:CMD}
\end{figure*}

\section{Identifications}
\label{ident}

We have compared our source list with the X-ray sources detected by \citet{SciortinoDamiani1998} in a {\it ROSAT} observation which includes field~2. We detected all the {\it ROSAT} sources which fall in the {\it XMM-Newton} field of view with an offset always smaller than $6\arcsec$, except for the USco~21 ROSAT source which is located $20\arcsec$ from the {\it XMM-Newton} counterpart (source~1 in Table~\ref{tab:src}). However this offset is compatible with the position uncertainty of the {\it ROSAT} PSPC point spread function.

We have searched for NIR counterparts to the X-ray sources in the 2MASS and in the DENIS catalogs. We have also searched for optical counterparts in the photometric CTIO and Danish 1.54m observations. All these catalogs and observations surveyed different sky areas, and only the 2MASS catalog covers entirely the two {\it XMM-Newton} fields. Fig.~\ref{fig:CATfov} shows the {\it XMM-Newton} fields together with the area covered by the DENIS, CTIO, and Danish 1.54m surveys.

Before assigning counterparts to the detected X-ray sources we have used the 2MASS catalog to check the astrometric accuracy of the {\it XMM-Newton} observations. We have measured for both observations a systematic and significant offset of $1.5\arcsec$ along RA, by comparing the position of X-ray and 2MASS sources. Hence we have corrected X-ray source positions to fix this offset. The astrometry of the CTIO and Danish 1.54m observations has also been registered on the 2MASS coordinate system.

We have adopted an identification radius between X-ray and optical/NIR sources of $5\arcsec$, estimated from the position offset histograms. This value is comparable to the FWHM of the point spread function of EPIC instruments.

Searching in the 2MASS point source catalog we have found counterparts for 60 X-ray sources, listed in Table~\ref{tab:src}. Only for source~1 we found two counterparts within $5\arcsec$. We have evaluated the expected numbers of spurious identifications assuming uncorrelated X-ray and optical/NIR source positions and uniform source distribution over the inspected sky area\footnote{Hence the derived estimates are upper limits.}. We have obtained that up to 6.5 and 9.5 identifications with 2MASS sources may be spurious for fields 1 and 2, respectively.

Only 71\% and 78\% of the two {\it XMM-Newton} fields are covered by the DENIS catalog (see Fig.~\ref{fig:CATfov}). We have assigned DENIS counterparts to 53 X-ray sources (out of 170), indicated in Table~\ref{tab:src}, and we found 2 counterparts only for source 171. We expect up to 12 and 6.7 spurious identifications for fields 1 and 2, respectively.

Among the optical sources detected in the CTIO observations (which surveyed the field~2) we have found counterparts for 11 X-ray sources (out of 107), with up to $1.6$ spurious identifications.

The Danish 1.54m observations cover 39\% of the {\it XMM-Newton} field~2. We have identified Danish counterparts for 13 X-ray sources (out of 54), with up to $3.2$ expected spurious identifications. Note that among Danish 1.54m sources we have found two possible counterparts for source~1, whose position coincides with the two 2MASS counterparts.

Totally we have found NIR or optical counterparts for 67 X-ray sources.

\subsection{Members of Upper Scorpius}
\label{uscomemb}

\begin{sidewaystable*}
\begin{minipage}[t]{\textwidth}
\renewcommand{\baselinestretch}{1.3}
\caption{Probable members of Upper Scorpius.}
\begin{center}
\label{tab:memb}
\scriptsize
\begin{tabular}{clrrrrrrrrrrrrrrllc}
\hline\hline
 \multicolumn{1}{c}{Source} & \multicolumn{1}{c}{Name} & \multicolumn{3}{c}{2MASS}                                                   & \multicolumn{3}{c}{DENIS}                                                   & \multicolumn{5}{c}{CTIO}                                                                                                        & \multicolumn{3}{c}{DANISH}                                                  & \multicolumn{1}{c}{$\log f_{\rm X}$}               & \multicolumn{1}{c}{$\log L_{\rm X}^{\rm a}$} & Memb$^{\rm b}$ \\
                            &                          & \multicolumn{1}{c}{$J$} & \multicolumn{1}{c}{$H$} & \multicolumn{1}{c}{$K$} & \multicolumn{1}{c}{$I$} & \multicolumn{1}{c}{$J$} & \multicolumn{1}{c}{$K$} & \multicolumn{1}{c}{$U$} & \multicolumn{1}{c}{$B$} & \multicolumn{1}{c}{$V$} & \multicolumn{1}{c}{$R$} & \multicolumn{1}{c}{$I$} & \multicolumn{1}{c}{$V$} & \multicolumn{1}{c}{$R$} & \multicolumn{1}{c}{$I$} & \multicolumn{1}{c}{(${\rm erg\,s^{-1}\,cm^{-2}}$)} & \multicolumn{1}{c}{${\rm (erg\,s^{-1})}$}    &                \\
\hline
              1 &            RX J155611.1-235054 &                12.19 &                11.56 &                11.30 &                13.91 &                12.22 &                11.24 &    $\cdot\cdot\cdot$ &    $\cdot\cdot\cdot$ &    $\cdot\cdot\cdot$ &    $\cdot\cdot\cdot$ &    $\cdot\cdot\cdot$ &                17.04 &                15.62 &                13.94 &      $\hspace{0.91em}    -13.27^{\rm c}$ &      $\hspace{0.91em}     29.13^{\rm c}$ &   Yes \\
              7 &                   V* V1144 Sco &                 9.66 &                 8.97 &                 8.74 &    $\cdot\cdot\cdot$ &    $\cdot\cdot\cdot$ &    $\cdot\cdot\cdot$ &                15.55 &                14.43 &                13.73 &                11.94 &                11.39 &                13.06 &                12.01 &                10.92 &      $\hspace{0.91em}    -12.05^{\rm d}$ &      $\hspace{0.91em}     30.36^{\rm d}$ &   W94 \\
             31 &                                &                13.47 &                12.85 &                12.53 &    $\cdot\cdot\cdot$ &    $\cdot\cdot\cdot$ &    $\cdot\cdot\cdot$ &    $\cdot\cdot\cdot$ &    $\cdot\cdot\cdot$ &    $\cdot\cdot\cdot$ &    $\cdot\cdot\cdot$ &    $\cdot\cdot\cdot$ &                18.85 &                17.30 &                15.39 &      $\hspace{0.91em}    -14.33^{\rm c}$ &      $\hspace{0.91em}     28.07^{\rm c}$ &   Yes \\
             32 &                                &                12.98 &                12.36 &                12.06 &    $\cdot\cdot\cdot$ &    $\cdot\cdot\cdot$ &    $\cdot\cdot\cdot$ &    $\cdot\cdot\cdot$ &    $\cdot\cdot\cdot$ &    $\cdot\cdot\cdot$ &    $\cdot\cdot\cdot$ &    $\cdot\cdot\cdot$ &                18.95 &                17.37 &                15.23 &      $\hspace{0.91em}    -13.67^{\rm d}$ &      $\hspace{0.91em}     28.73^{\rm d}$ &   Yes \\
             39 &                                &                13.82 &                13.19 &                12.84 &                15.53 &                13.71 &                12.86 &    $\cdot\cdot\cdot$ &    $\cdot\cdot\cdot$ &    $\cdot\cdot\cdot$ &    $\cdot\cdot\cdot$ &    $\cdot\cdot\cdot$ &    $\cdot\cdot\cdot$ &    $\cdot\cdot\cdot$ &    $\cdot\cdot\cdot$ &      $\hspace{0.91em}    -13.93^{\rm c}$ &      $\hspace{0.91em}     28.47^{\rm c}$ &    ?  \\
             42 &            RX J155627.5-233848 &                10.09 &                 9.39 &                 9.11 &    $\cdot\cdot\cdot$ &    $\cdot\cdot\cdot$ &    $\cdot\cdot\cdot$ &                17.32 &                15.73 &                14.14 &                12.99 &                11.59 &                14.28 &                13.04 &                11.67 &      $\hspace{0.91em}    -12.90^{\rm d}$ &      $\hspace{0.91em}     29.50^{\rm d}$ &   Yes \\
             43 &                   V* V1146 Sco &                 9.73 &                 9.07 &                 8.88 &                10.95 &                 9.77 &                 8.85 &                15.18 &                14.00 &                12.63 &                11.78 &                10.86 &    $\cdot\cdot\cdot$ &    $\cdot\cdot\cdot$ &    $\cdot\cdot\cdot$ &      $\hspace{0.91em}    -12.59^{\rm d}$ &      $\hspace{0.91em}     29.81^{\rm d}$ &   W94 \\
             45 &       DENIS-P J155601.0-233808 &                13.86 &                13.24 &                12.81 &                16.33 &                13.96 &                12.85 &    $\cdot\cdot\cdot$ &    $\cdot\cdot\cdot$ &    $\cdot\cdot\cdot$ &    $\cdot\cdot\cdot$ &    $\cdot\cdot\cdot$ &    $\cdot\cdot\cdot$ &    $\cdot\cdot\cdot$ &    $\cdot\cdot\cdot$ &      $\hspace{0.91em}    -14.30^{\rm c}$ &      $\hspace{0.91em}     28.11^{\rm c}$ &   A00 \\
             62 &            RX J155620.6-233606 &                11.26 &                10.57 &                10.32 &                12.81 &                11.32 &                10.26 &                17.83 &                17.27 &                15.50 &                14.33 &                12.82 &                15.67 &                14.37 &                12.89 &      $\hspace{0.91em}    -13.31^{\rm d}$ &      $\hspace{0.91em}     29.10^{\rm d}$ &   Yes \\
             66 &                                &                13.26 &                12.60 &                12.30 &                15.15 &                13.28 &                12.21 &    $\cdot\cdot\cdot$ &    $\cdot\cdot\cdot$ &    $\cdot\cdot\cdot$ &    $\cdot\cdot\cdot$ &    $\cdot\cdot\cdot$ &                18.32 &                16.94 &                15.19 &      $\hspace{0.91em}    -13.96^{\rm c}$ &      $\hspace{0.91em}     28.45^{\rm c}$ &   Yes \\
             70 &                                &                13.93 &                13.37 &                13.04 &                15.50 &                13.88 &                13.00 &    $\cdot\cdot\cdot$ &    $\cdot\cdot\cdot$ &    $\cdot\cdot\cdot$ &    $\cdot\cdot\cdot$ &    $\cdot\cdot\cdot$ &    $\cdot\cdot\cdot$ &    $\cdot\cdot\cdot$ &    $\cdot\cdot\cdot$ &      $\hspace{0.91em}    -14.26^{\rm c}$ &      $\hspace{0.91em}     28.14^{\rm c}$ &    ?  \\
             95 &                   V* V1145 Sco &                11.22 &                10.58 &                10.28 &                12.80 &                11.31 &                10.33 &                17.65 &                16.73 &                15.28 &                14.16 &                12.69 &    $\cdot\cdot\cdot$ &    $\cdot\cdot\cdot$ &    $\cdot\cdot\cdot$ &      $\hspace{0.91em}    -12.66^{\rm d}$ &      $\hspace{0.91em}     29.75^{\rm d}$ &   W94 \\
            100 &                      HD 142578 &                 8.05 &                 7.93 &                 7.84 &                 8.74 &                 8.04 &                 7.87 &    $\cdot\cdot\cdot$ &    $\cdot\cdot\cdot$ &    $\cdot\cdot\cdot$ &    $\cdot\cdot\cdot$ &    $\cdot\cdot\cdot$ &    $\cdot\cdot\cdot$ &    $\cdot\cdot\cdot$ &    $\cdot\cdot\cdot$ &      $\hspace{0.91em}    -11.07^{\rm d}$ &      $\hspace{0.91em}     31.33^{\rm d}$ &   S92 \\
            114 &                GSC 06793-00569 &                 9.32 &                 8.62 &                 8.49 &                10.28 &                 9.31 &                 8.43 &    $\cdot\cdot\cdot$ &    $\cdot\cdot\cdot$ &    $\cdot\cdot\cdot$ &    $\cdot\cdot\cdot$ &    $\cdot\cdot\cdot$ &    $\cdot\cdot\cdot$ &    $\cdot\cdot\cdot$ &    $\cdot\cdot\cdot$ &      $\hspace{0.91em}    -12.07^{\rm d}$ &      $\hspace{0.91em}     30.33^{\rm d}$ &   P98 \\
            123 &                                &                12.54 &                11.92 &                11.64 &    $\cdot\cdot\cdot$ &    $\cdot\cdot\cdot$ &    $\cdot\cdot\cdot$ &    $\cdot\cdot\cdot$ &    $\cdot\cdot\cdot$ &    $\cdot\cdot\cdot$ &    $\cdot\cdot\cdot$ &    $\cdot\cdot\cdot$ &    $\cdot\cdot\cdot$ &    $\cdot\cdot\cdot$ &    $\cdot\cdot\cdot$ &      $\hspace{0.91em}    -13.95^{\rm c}$ &      $\hspace{0.91em}     28.46^{\rm c}$ &   Yes \\
            125 &                                &                10.57 &                 9.88 &                 9.60 &    $\cdot\cdot\cdot$ &    $\cdot\cdot\cdot$ &    $\cdot\cdot\cdot$ &    $\cdot\cdot\cdot$ &    $\cdot\cdot\cdot$ &    $\cdot\cdot\cdot$ &    $\cdot\cdot\cdot$ &    $\cdot\cdot\cdot$ &    $\cdot\cdot\cdot$ &    $\cdot\cdot\cdot$ &    $\cdot\cdot\cdot$ &      $\hspace{0.91em}    -12.51^{\rm d}$ &      $\hspace{0.91em}     29.90^{\rm d}$ &   Yes \\
            132 &                GSC 06793-00819 &                 8.28 &                 7.71 &                 7.46 &                 9.34 &                 8.34 &                 7.56 &    $\cdot\cdot\cdot$ &    $\cdot\cdot\cdot$ &    $\cdot\cdot\cdot$ &    $\cdot\cdot\cdot$ &    $\cdot\cdot\cdot$ &    $\cdot\cdot\cdot$ &    $\cdot\cdot\cdot$ &    $\cdot\cdot\cdot$ &      $\hspace{0.91em}    -11.58^{\rm d}$ &      $\hspace{0.91em}     30.83^{\rm d}$ &   P98 \\
            144 &                                &                12.71 &                12.09 &                11.76 &                14.77 &                12.74 &                11.76 &    $\cdot\cdot\cdot$ &    $\cdot\cdot\cdot$ &    $\cdot\cdot\cdot$ &    $\cdot\cdot\cdot$ &    $\cdot\cdot\cdot$ &    $\cdot\cdot\cdot$ &    $\cdot\cdot\cdot$ &    $\cdot\cdot\cdot$ &      $\hspace{0.91em}    -13.99^{\rm c}$ &      $\hspace{0.91em}     28.41^{\rm c}$ &   Yes \\
            156 &                                &                13.76 &                13.12 &                12.79 &                16.07 &                13.80 &                12.85 &    $\cdot\cdot\cdot$ &    $\cdot\cdot\cdot$ &    $\cdot\cdot\cdot$ &    $\cdot\cdot\cdot$ &    $\cdot\cdot\cdot$ &    $\cdot\cdot\cdot$ &    $\cdot\cdot\cdot$ &    $\cdot\cdot\cdot$ &      $\hspace{0.91em}    -14.19^{\rm c}$ &      $\hspace{0.91em}     28.22^{\rm c}$ &   Yes \\
            158 &                GSC 06793-00994 &                 9.38 &                 8.77 &                 8.61 &                10.24 &                 9.31 &                 8.56 &    $\cdot\cdot\cdot$ &    $\cdot\cdot\cdot$ &    $\cdot\cdot\cdot$ &    $\cdot\cdot\cdot$ &    $\cdot\cdot\cdot$ &    $\cdot\cdot\cdot$ &    $\cdot\cdot\cdot$ &    $\cdot\cdot\cdot$ &      $\hspace{0.91em}    -11.88^{\rm d}$ &      $\hspace{0.91em}     30.52^{\rm d}$ &   P98 \\
            173 &                                &                10.60 &                 9.81 &                 9.59 &                11.97 &                10.53 &                 9.49 &    $\cdot\cdot\cdot$ &    $\cdot\cdot\cdot$ &    $\cdot\cdot\cdot$ &    $\cdot\cdot\cdot$ &    $\cdot\cdot\cdot$ &    $\cdot\cdot\cdot$ &    $\cdot\cdot\cdot$ &    $\cdot\cdot\cdot$ &      $\hspace{0.91em}    -12.23^{\rm d}$ &      $\hspace{0.91em}     30.17^{\rm d}$ &   Yes \\
            180 &                GSC 06793-00797 &                 9.32 &                 8.65 &                 8.45 &                10.30 &                 9.26 &                 8.38 &    $\cdot\cdot\cdot$ &    $\cdot\cdot\cdot$ &    $\cdot\cdot\cdot$ &    $\cdot\cdot\cdot$ &    $\cdot\cdot\cdot$ &    $\cdot\cdot\cdot$ &    $\cdot\cdot\cdot$ &    $\cdot\cdot\cdot$ &      $\hspace{0.91em}    -11.78^{\rm d}$ &      $\hspace{0.91em}     30.63^{\rm d}$ &   P98 \\
\hline 
                &                    UScoCTIO 92 &                13.49 &                12.90 &                12.54 &                15.48 &                13.52 &                12.49 &    $\cdot\cdot\cdot$ &    $\cdot\cdot\cdot$ &    $\cdot\cdot\cdot$ &    $\cdot\cdot\cdot$ &    $\cdot\cdot\cdot$ &    $\cdot\cdot\cdot$ &    $\cdot\cdot\cdot$ &    $\cdot\cdot\cdot$ &      $              <    -14.21^{\rm c}$ &      $              <     28.19^{\rm c}$ &   A00 \\
                &                   UScoCTIO 104 &                13.48 &                12.86 &                12.59 &                15.70 &                13.47 &                12.53 &    $\cdot\cdot\cdot$ &    $\cdot\cdot\cdot$ &    $\cdot\cdot\cdot$ &    $\cdot\cdot\cdot$ &    $\cdot\cdot\cdot$ &    $\cdot\cdot\cdot$ &    $\cdot\cdot\cdot$ &    $\cdot\cdot\cdot$ &      $              <    -14.18^{\rm c}$ &      $              <     28.22^{\rm c}$ &   A00 \\
                &                   UScoCTIO 137 &                15.66 &                15.00 &                14.42 &    $\cdot\cdot\cdot$ &    $\cdot\cdot\cdot$ &    $\cdot\cdot\cdot$ &    $\cdot\cdot\cdot$ &    $\cdot\cdot\cdot$ &    $\cdot\cdot\cdot$ &    $\cdot\cdot\cdot$ &    $\cdot\cdot\cdot$ &    $\cdot\cdot\cdot$ &    $\cdot\cdot\cdot$ &    $\cdot\cdot\cdot$ &      $              <    -14.31^{\rm c}$ &      $              <     28.09^{\rm c}$ &   A00 \\
                &        2MASS J16131211-2305031 &                14.05 &                13.45 &                13.01 &                16.64 &                14.17 &                13.03 &    $\cdot\cdot\cdot$ &    $\cdot\cdot\cdot$ &    $\cdot\cdot\cdot$ &    $\cdot\cdot\cdot$ &    $\cdot\cdot\cdot$ &    $\cdot\cdot\cdot$ &    $\cdot\cdot\cdot$ &    $\cdot\cdot\cdot$ &      $              <    -14.34^{\rm c}$ &      $              <     28.06^{\rm c}$ &   S06 \\
\hline
\end{tabular}
\normalsize
\end{center}
 
$^{\rm a}$~X-ray luminosities were evaluated assuming a distance of 145\,pc.  \\
$^{\rm b}$~Question mark indicates stars whose photometry casts some doubt on their membership of Upper Scorpius. W94, A00, S92, P98, S06 designate stars that were identifyed as Upper Scorpius members by \citet{WalterVrba1994}, \citet{ArdilaMartin2000}, \citet{SlawsonHill1992}, \citet{PreibischGuenther1998}, and \citet{SlesnickCarpenter2006} respectively. \\
$^{\rm c}$~Unabsorbed X-ray fluxes and luminosities, in the $0.5-8.0$\,keV band, were computed starting from PN equivalent count rates and adopting a multiplicative factor of $2.37\times10^{-12}\,{\rm erg\,cm^{-2}}$.  \\
$^{\rm d}$~Unabsorbed X-ray fluxes and luminosities, in the $0.5-8.0$\,keV band, were evaluated from the best fit model of the observed spectrum. \\
\vfill
\end{minipage}
\end{sidewaystable*}

In order to establish which X-ray sources are likely USco members we have inspected the color magnitude diagrams (CMD) of the counterparts, displayed in Fig.~\ref{fig:CMD}.

The USco association covers on the sky an area of $\sim150\,{\rm deg^{2}}$, with a projected diameter of $\sim14\,{\rm deg}$, which corresponds to $\sim35$\,pc at the distance of the association \citep{PreibischBrown2002}. Assuming a spherical distribution, the distance spread of USco members in the radial direction should also be $\sim35$\,pc. We have delineated with light gray in the CMDs of Fig.~\ref{fig:CMD} the region occupied by the isochrones if the distance is changed by $\pm20$\,pc.

We have selected as photometric USco members sources whose photometry is compatible with the values predicted by \cite{SiessDufour2000} or \cite{BaraffeChabrier1998}. We have identified 22 photometric USco members among the 224 detected X-ray sources. In each CMD gray dots mark X-ray undetected sources, black diamonds X-ray detected stars that we consider as photometric USco members, black crosses sources with X-ray counterparts whose photometry is not compatible with USco membership.

All the X-ray sources identified as photometric USco members are detected in the 2MASS survey. Note that the 2MASS catalog contains objects much fainter than the low-mass stars and brown dwarfs of the USco association, as suggested from upper left panel of Fig.~\ref{fig:CMD}.

Members are listed in Table~\ref{tab:memb} together with NIR and optical photometry, and available SIMBAD counterparts. The X-ray sources selected as photometric USco members are also indicated with circles in Fig.~\ref{fig:XMMfov}. As indicated in Table~\ref{tab:memb} only 9 of the 22 selected stars were already known USco members.

In some uncertain cases, like the two X-ray sources 39 and 70 whose DENIS photometry is located significantly below both the isochrones (upper right panel of Fig.~\ref{fig:CMD}), we have decided on the basis of the 2MASS data. However we have labeled these sources as uncertain USco members in Table~\ref{tab:memb}.

As mentioned above, X-ray source~1 is associated with two counterparts, located at $1.5$ and $4\,\arcsec$, detected both in the 2MASS and Danish 1.54m surveys: the first is the source 2MASS~J15561238$-$2351030 (\#1193 in Table~C.1); the second is the source 2MASS~J15561210$-$2351065 (\#1189 in Table~C.1). In both surveys the magnitudes of the former counterpart are compatible with USco membership (see Table~\ref{tab:memb}). For the latter counterpart the 2MASS photometry is $J=16.04$, $H=15.20$, and $K>13.09$, which suggests it to be a background object, while its magnitudes derived from the Danish 1.54m observation are likely contaminated by the former counterpart. Since X-ray emission is more plausible from a young and nearby star than from a background object, and because the offset is smaller, we have assumed that the counterpart of the X-ray source~1 is the source 2MASS~J15561238-2351030, which we have selected as photometric USco member.

The class of the detected X-ray USco sources ranges from G to late M. The only early star is the counterpart of the X-ray source~100. The set of selected photometric members contains also a known brown dwarf (X-ray source 45).

The fields under investigation contain four other known USco members, which were not detected in the X-ray observations: UScoCTIO~92, UScoCTIO~104, UScoCTIO~137, and 2MASS~J16131211-2305031 \citep{ArdilaMartin2000,SlesnickCarpenter2006}, with the latter three being brown dwarfs \citep{ArdilaMartin2000,JayawardhanaArdila2003,SlesnickCarpenter2006}. We list them in the lower part of Table~\ref{tab:memb} indicating the upper limits to their X-ray emission.

Considering the vicinity of the USco association (145\,pc) and the low extinction, it is likely that none of the X-ray sources without optical/NIR counterparts is an USco member.

\subsubsection{Comments on individual USco members}
\label{indstars}

Source 1 is located $20\,\arcsec$ away from the X-ray source ${\rm RX~J155611.1-235054}$ (USco~21) detected in a {\it ROSAT} observation by \citet{SciortinoDamiani1998}. Given the uncertainty on {\it ROSAT} positions the two sources likely coincide, and therefore we assume them to be the same source.

Source 7 (USco~27, ScoPMS~13) is a M1.5 WTTS, it is a binary system with the two companions separated by $0\farcs09$ \citep{KohlerKunkel2000}. Its young age was shown by \citet{WalterVrba1994} and \citet{Martin1998} who measured the H$\alpha$ emission and high Li abundance. \citet{AdamsWalter1998} derived a rotation period of $2.5\pm0.3$\,d indicating that this star is a rapid rotator with a period compatible with the average rotation period of WTTSs. 

Spectroscopy of source~43 (USco~34, ScoPMS~15) revealed a high lithium abundance and H$\alpha$ emission which place this source among the WTTSs belonging to the USco association \citep{WalterVrba1994,Martin1998}. Source~43, whose spectral type is M0, displays a rotational period of $6.2\pm0.4$\,d \citep{AdamsWalter1998} which, combined with the $v\sin i$ measure of $18\,{\rm km\,s^{-1}}$ and the predicted radius of $2\,R_{\sun}$, indicates an inclination angle $i$ of $\sim90\deg$.

Source~45 (UScoCTIO~113, DENIS-P J155601.0-233808) was assigned to the USco association by \citet{ArdilaMartin2000} which selected it for its photometric properties. \citet{MartinDelfosse2004} performed optical spectroscopy measuring an H$\alpha$ in emission with an equivalent width of $-20$\,\AA, and deriving a spectral type of M6.5. These finding suggests this source to be a non-accreting brown dwarf. The X-ray emission from this source was already reported by \citet{Bouy2004} based on the same {\it XMM-Newton} observation we present in this work.

\citet{WalterVrba1994} showed that source~95 (USco~21, ScoPMS~14) is an M3 WTTS whose photometry is compatible with USco membership. Its rotation period of 2.2\,d was measured by \citet{AdamsWalter1998}.

We have identified source~100 with the intermediate mass star HD~142578, whose spectral type listed in the SIMBAD database is A2. Photometric measures were conducted by \citet{SlawsonHill1992}. Since neither coronal nor wind X-ray emission is expected from this type of stars, it is possible that the X-ray radiation is due to an unseen low-mass companion \citep[see][and references therein]{StelzerFlaccomio2005}.

Source~114, 158 and 180 are WTTSs of spectral type K1, G4, and K4, respectively, which show large Li abundance and weak H$\alpha$ emission \citep{PreibischGuenther1998}.

\citet{PreibischGuenther1998} identified source~132 (GSC~06793-00819) as a K0 USco member. They performed optical spectroscopy of this star and measured high Li abundance and H$\alpha$ line in absorption with ${\rm EW=0.96}$\,\AA. Asymmetric features in the H$\alpha$ were observed by \citet{MamajekMeye2004} who also registered an excess in the $N$-band. From these findings \citeauthor{MamajekMeye2004} proposed this star to be actively accreting from its circumstellar disk.

\begin{figure}[t]
\resizebox{\hsize}{!}{\includegraphics{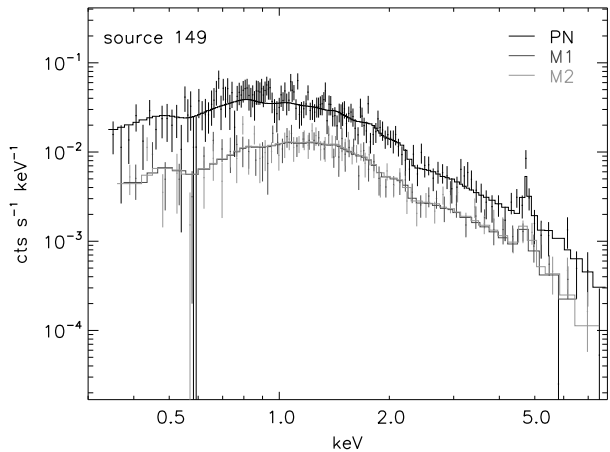}}
\caption{Observed and best fit EPIC spectra of the intra cluster medium of the newly discovered galaxy cluster (source 149).}
\label{fig:icmspec}
\end{figure}

\subsection{Extragalactic X-ray sources}

In the two {\it XMM-Newton} observations ($\sim0.4\,\deg^2$) we have detected 224 sources, and only 22 have been selected as photometric USco members.

The on-axis detection limits for extragalactic X-ray sources are $3.0$ and $3.3\times10^{-15}\,{\rm erg\,s^{-1}\,cm^{-2}}$ for fields 1 and 2, assuming a power law spectrum with index $1.6$. The sensitivity however decreases significantly for increasing off-axis angles. Considering the extragalactic X-ray population in the $0.5-8.0$\,keV band \citep{AlexanderBauer2003}, the expected number of extragalactic X-ray sources in the two analyzed {\it XMM/Newton} observations is $\sim103$ and $\sim95$. Hence only very few X-ray galactic sources, other than the USco members, should be present in the inspected sky areas.

Source~149, near the telescope axis in the {\it XMM-Newton} field~1 observation (see Fig.~\ref{fig:XMMfov}), shows an extended X-ray emission, which resembles the typical X-ray emission of the intra cluster medium of a galaxy cluster \citep[e.g.][and references therein]{RosatiBorgani2002}. No bibliographic information is available for this source. We have fitted its EPIC spectra, extracted within $60\arcsec$ from the source position, adopting as model an absorbed optically thin isothermal plasma (see Fig.~\ref{fig:icmspec}). The best fit model indicates a redshift of $0.41\pm0.02$ (evident from the displacement of the \ion{Fe}{xxiv} line from 6.7 to 4.7\,keV), while the best fit temperature is $60\pm10$\,MK, and the metallicity is $0.4\pm0.2$ with respect to the solar photospheric value.


\section{Spectral analysis}

In this section we report the X-ray spectral analysis of photometric USco members performed without distinguishing between flaring and quiescent emission. Therefore our analysis represent a study of the time averaged properties of PMS X-ray emission on time scales of $\sim40-50$\,ks.

The spectral analysis has been performed using XSPEC V11.3.0, and adopting the Astrophysical Plasma Emission Database \citep[APED V1.3,][]{SmithBrickhouse2001}, based on the ionization equilibrium from \citet{MazzottaMazzitelli1998}.

\begin{figure}[t]
\resizebox{\hsize}{!}{\includegraphics{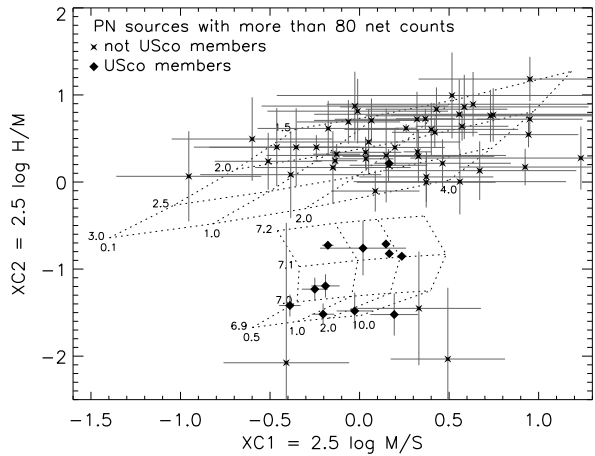}}
\caption{X-ray colors of PN sources with more than 80 background subtracted counts. $S$, $M$ and $H$ indicate the number of net counts in the soft ($0.3-0.85$\,keV), medium ($0.85-1.4$\,keV) and hard ($1.4-7.9$\,keV) energy bands. The upper grid refers to power law spectra with indices ranging between 1.5 and 3.0 (indicated in the left part of the grid), and hydrogen column density ranging between 0.1 and $4\times10^{21}\,{\rm cm^{-2}}$ (indicated in the lower part of the grid). The lower grid corresponds to thermal optically thin plasma with two thermal components: the lower temperature is kept constant at $\log T_{L} {\rm (K)}=6.4$, and the higher temperature ranges between $\log T_{H} {\rm (K)}=6.9$ and 7.2 (indicated on the left part of the grid); the emission measure ratio between the two components ${\it EM}_{H}/{\it EM}_{C}$ varies from 0.5 to 10 (indicated in the lower part of the grid); the hydrogen column density and the metallicity are $1.5\times10^{21}\,{\rm cm^{-2}}$ and $0.15\,{\rm Fe_{\sun}}$, respectively (these values have been derived from the X-ray spectral analysis of the X-ray brightest USco sources, see Sect.~\ref{specfit}).}
\label{fig:obsxcol}
\end{figure}

\subsection{X-ray colors}
\label{xraycol}

\begin{figure*}
\centering
\includegraphics[width=17cm,height=22cm]{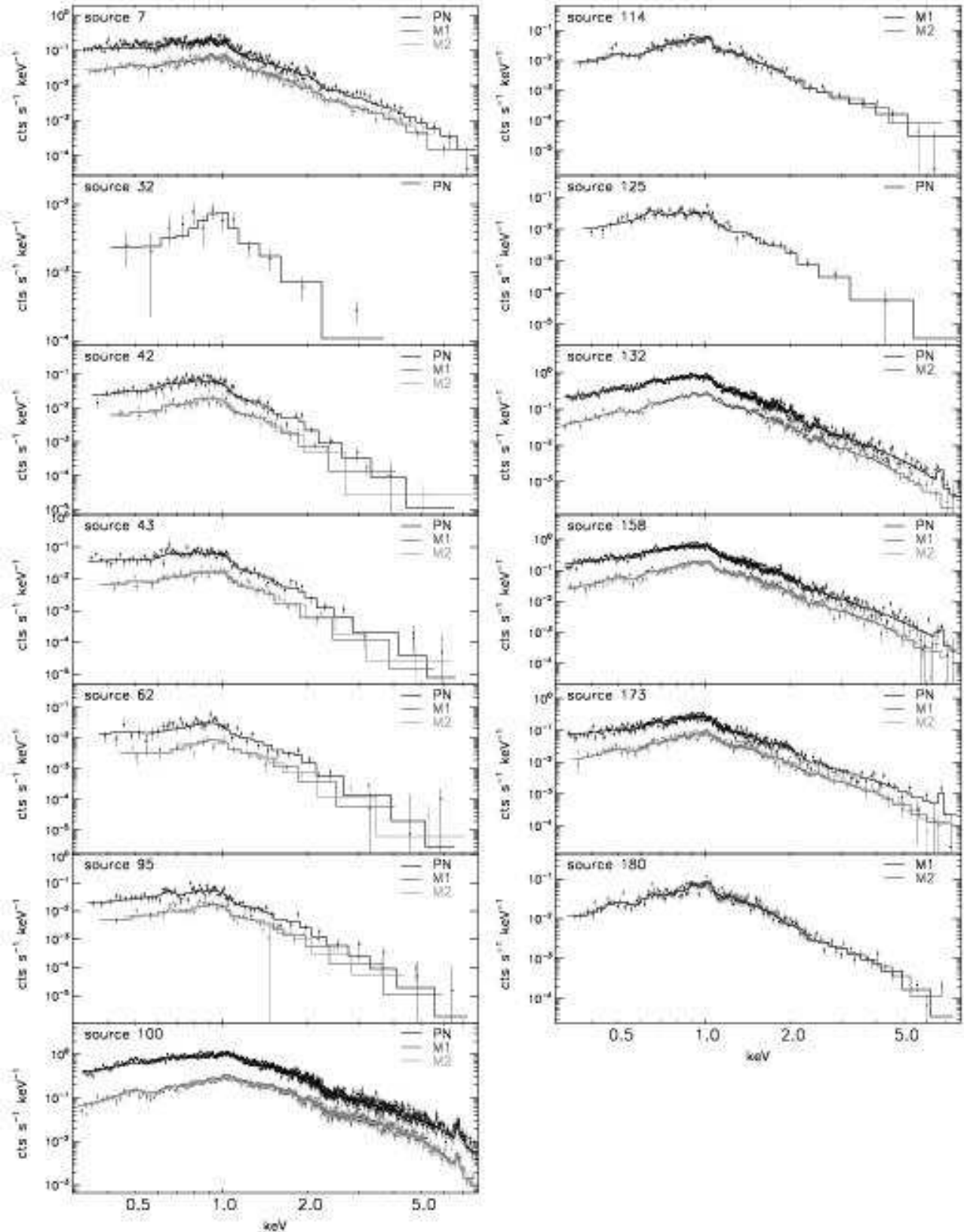}
\caption{EPIC spectra of the X-ray brightest USco members. Black indicates PN observed and predicted spectra, while dark gray and light gray mark the MOS1 and MOS2 spectra, respectively. Left column contains sources detected in the field~2 observation, while right column those of field~1.}
\label{fig:epicspec}
\end{figure*}

To discriminate among different classes of detected X-ray sources we have computed two X-ray colors. We have divided the EPIC energy band into three ranges: soft $0.3-0.85$, medium $0.85-1.4$, and hard $1.4-7.9$\,keV. For each X-ray source we have evaluated the background subtracted number of photons in the three bands, indicating them with $S$, $M$ and $H$. Then we have computed the X-ray colors as ${\it XC1}=2.5\log(M/S)$ and ${\it XC2}=2.5\log(H/M)$, which are listed in Table~\ref{tab:src}.

In Fig.~\ref{fig:obsxcol} we show the observed X-ray colors for the strongest PN sources. We have compared the observed values with predictions based on two types of models: power law spectra and optically thin collisionally excited plasma, represented in Fig.~\ref{fig:obsxcol} by the upper and lower dotted grids.

All the USco members have {\it XC1} and {\it XC2} colors well compatible with the expected values for coronal sources. The only USco member in this plot which falls in the upper grid is source~100, the intermediate mass star HD~142578, which has ${\it XC2}=0.20\pm0.01$. This source is characterized by a large flare (see Sect.~\ref{source100flare}) during which the count rate increases by a factor $\sim100$ and the plasma temperature reaches $T\sim70$\,MK. However the {\it XC2} color of its quiet phase is $-1.14\pm0.18$ and hence compatible with the other USco members.

The observed {\it XC2} values have bimodal distribution separated by the discriminating value ${\it XC2}\approx-0.5$. USco members, except source 100, have ${\it XC2}<-0.5$. Three other X-ray sources (sources 182, 200 and 209), display {\it XC2} colors compatible with, or even softer than, that of the USco members. These three sources have 2MASS counterparts, and their X-ray colors suggest them to be background stars. On the other hand, most non USco members in Fig.~\ref{fig:obsxcol} show positive {\it XC2} values. Only 20\% of these X-ray sources have 2MASS counterparts, compatible with most of them being extragalactic sources with harder spectra.

The {\it XC1} color is very sensitive to the column density value. A change in the $N_{\rm H}$ produces a variation in the {\it XC1} colors which is twice the corresponding variation of {\it XC2}. In particular a change of a factor 2 for $N_{\rm H}$ causes an increase of {\it XC1} of $\sim0.4$. Since the predicted grids, evaluated for typical coronal temperatures, describe well the observed X-ray colors of selected USco members we are confident that the adopted $N_{\rm H}$ is reliable.

\subsection{Spectral fitting}
\label{specfit}

We have analyzed the X-ray spectra of all the sources identified as photometric members of the USco association with more than 200 total PN counts. We have also analyzed the spectra of the two USco stars (source 114 and 180) which fall outside the PN detector but with high $S/N$ MOS spectra. With these criteria we have selected a subset of 13 sources among the 22 photometric members. We have performed simultaneous spectral fitting of PN, MOS1 and MOS2 data, except for the faintest sources where only the PN spectrum has been studied. Finally, for source 132 we have not considered MOS1 data since they were not compatible with PN and MOS2 data\footnote{A column of bad pixels of MOS1, located exactly on the source position, likely affects the effective area estimation.}.

\begin{sidewaystable*}
\begin{minipage}[t]{\textwidth}
\renewcommand{\baselinestretch}{1.3}
\caption{Fit results of the EPIC spectra of the Upper Scorpius members.}
\begin{center}
\label{tab:fitres}
\normalsize
\begin{tabular}{ccccccccccccc}
\hline\hline
 Source & Instrument$^{\rm a}$ & $N_{\rm H}$               & $\log T_{1}$ & $\log T_{2}$ & $\log T_{3}$ & $\log {\it EM}_{1}$ & $\log {\it EM}_{2}$ & $\log {\it EM}_{3}$ & Fe$^{\rm b}$ & O$^{\rm b}$ & Ne$^{\rm b}$ & $\chi^2/{\rm dof}$ \\
        &                      & ${\rm(10^{21}\,cm^{-2})}$ & (K)          & (K)          & (K)          & ${\rm(cm^{-3})}$    & ${\rm(cm^{-3})}$    & ${\rm(cm^{-3})}$    &              &             &              &                    \\
\hline
   7 &   PN+M1+M2 &      $ 0.32^{+ 0.27}_{- 0.25}$ &      $ 6.64^{+ 0.05}_{- 0.05}$ &      $ 7.39^{+ 0.07}_{- 0.06}$ &              $\cdot\cdot\cdot$ &      $52.98^{+ 0.34}_{- 0.31}$ &      $53.26^{+ 0.06}_{- 0.09}$ &              $\cdot\cdot\cdot$ &      $ 0.20^{+ 0.21}_{- 0.11}$ &                   $= {\rm Fe}$ &      $ 1.12^{+ 1.11}_{- 0.57}$ &     $     438.2 /       391$ \\
  32 &         PN &                       $= 1.50$ &      $ 7.06^{+ 0.10}_{- 0.10}$ &              $\cdot\cdot\cdot$ &              $\cdot\cdot\cdot$ &      $51.93^{+ 0.06}_{- 0.07}$ &              $\cdot\cdot\cdot$ &              $\cdot\cdot\cdot$ &                       $= 0.15$ &                   $= {\rm Fe}$ &                   $= {\rm Fe}$ &     $       8.5 /        11$ \\
  42 &   PN+M1+M2 &      $ 1.38^{+ 0.89}_{- 0.72}$ &      $ 6.55^{+ 0.14}_{- 0.17}$ &      $ 6.99^{+ 0.09}_{- 0.07}$ &              $\cdot\cdot\cdot$ &      $52.65^{+ 0.40}_{- 0.43}$ &      $52.51^{+ 0.20}_{- 0.25}$ &              $\cdot\cdot\cdot$ &      $ 0.14^{+ 0.14}_{- 0.05}$ &                   $= {\rm Fe}$ &                   $= {\rm Fe}$ &     $     116.2 /       120$ \\
  43 &   PN+M1+M2 &      $ 0.55^{+ 0.65}_{- 0.50}$ &      $ 6.60^{+ 0.16}_{- 0.09}$ &      $ 7.05^{+ 0.05}_{- 0.07}$ &              $\cdot\cdot\cdot$ &      $52.79^{+ 0.37}_{- 0.34}$ &      $52.82^{+ 0.20}_{- 0.18}$ &              $\cdot\cdot\cdot$ &      $ 0.16^{+ 0.12}_{- 0.08}$ &                   $= {\rm Fe}$ &                   $= {\rm Fe}$ &     $     132.7 /       126$ \\
  62 &   PN+M1+M2 &      $ 0.88^{+ 0.52}_{- 0.47}$ &      $ 6.96^{+ 0.04}_{- 0.05}$ &              $\cdot\cdot\cdot$ &              $\cdot\cdot\cdot$ &      $52.42^{+ 0.14}_{- 0.15}$ &              $\cdot\cdot\cdot$ &              $\cdot\cdot\cdot$ &      $ 0.08^{+ 0.05}_{- 0.03}$ &                   $= {\rm Fe}$ &                   $= {\rm Fe}$ &     $      87.0 /        70$ \\
  95 &   PN+M1+M2 &      $ 1.39^{+ 1.25}_{- 0.69}$ &      $ 6.48^{+ 0.11}_{- 0.14}$ &      $ 6.98^{+ 0.07}_{- 0.06}$ &              $\cdot\cdot\cdot$ &      $52.88^{+ 0.48}_{- 0.49}$ &      $52.79^{+ 0.18}_{- 0.22}$ &              $\cdot\cdot\cdot$ &      $ 0.14^{+ 0.10}_{- 0.05}$ &                   $= {\rm Fe}$ &                   $= {\rm Fe}$ &     $     138.4 /       110$ \\
 100 &   PN+M1+M2 &      $ 1.30^{+ 0.44}_{- 0.25}$ &      $ 6.46^{+ 0.13}_{- 0.07}$ &      $ 7.24^{+ 0.06}_{- 0.07}$ &      $ 7.91^{+ 0.16}_{- 0.09}$ &      $53.72^{+ 0.55}_{- 0.51}$ &      $53.76^{+ 0.17}_{- 0.19}$ &      $54.01^{+ 0.06}_{- 0.07}$ &      $ 0.27^{+ 0.09}_{- 0.10}$ &      $ 0.13^{+ 0.24}_{- 0.08}$ &      $ 0.48^{+ 0.84}_{- 0.31}$ &     $    1494.8 /      1218$ \\
 114 &      M1+M2 &      $ 1.97^{+ 0.66}_{- 1.33}$ &      $ 6.72^{+ 0.21}_{- 0.07}$ &      $ 7.48^{+ 0.72}_{- 0.47}$ &              $\cdot\cdot\cdot$ &      $53.78^{+ 0.28}_{- 0.91}$ &      $52.62^{+ 0.67}_{- 0.49}$ &              $\cdot\cdot\cdot$ &      $ 0.05^{+ 0.15}_{- 0.02}$ &                   $= {\rm Fe}$ &      $ 0.34^{+ 0.52}_{- 0.16}$ &     $     149.4 /       120$ \\
 125 &         PN &      $ 2.18^{+ 2.20}_{- 1.32}$ &      $ 6.51^{+ 0.11}_{- 0.11}$ &      $ 7.09^{+ 0.07}_{- 0.11}$ &              $\cdot\cdot\cdot$ &      $53.02^{+ 0.76}_{- 0.68}$ &      $52.60^{+ 0.28}_{- 0.54}$ &              $\cdot\cdot\cdot$ &      $ 0.27^{+ 1.47}_{- 0.16}$ &                   $= {\rm Fe}$ &                   $= {\rm Fe}$ &     $      45.5 /        50$ \\
 132 &      PN+M2 &      $ 1.56^{+ 0.66}_{- 0.31}$ &      $ 6.58^{+ 0.05}_{- 0.03}$ &      $ 7.05^{+ 0.13}_{- 0.13}$ &      $ 7.42^{+ 1.54}_{- 0.09}$ &      $53.61^{+ 0.58}_{- 0.42}$ &      $53.44^{+ 0.40}_{- 0.36}$ &      $53.47^{+ 0.15}_{- 0.61}$ &      $ 0.21^{+ 0.13}_{- 0.13}$ &                   $= {\rm Fe}$ &      $ 0.46^{+ 0.50}_{- 0.26}$ &     $     633.0 /       602$ \\
 158 &   PN+M1+M2 &      $ 1.49^{+ 0.53}_{- 0.24}$ &      $ 6.61^{+ 0.05}_{- 0.04}$ &      $ 7.04^{+ 0.06}_{- 0.07}$ &      $ 7.38^{+ 0.41}_{- 0.07}$ &      $53.26^{+ 0.46}_{- 0.28}$ &      $53.16^{+ 0.36}_{- 0.27}$ &      $53.16^{+ 0.13}_{- 0.54}$ &      $ 0.22^{+ 0.11}_{- 0.12}$ &                   $= {\rm Fe}$ &      $ 0.50^{+ 0.32}_{- 0.24}$ &     $     766.0 /       670$ \\
 173 &   PN+M1+M2 &      $ 1.23^{+ 0.53}_{- 0.34}$ &      $ 6.62^{+ 0.13}_{- 0.08}$ &      $ 7.05^{+ 0.03}_{- 0.04}$ &      $ 7.46^{+ 0.28}_{- 0.10}$ &      $52.60^{+ 0.43}_{- 0.33}$ &      $52.78^{+ 0.34}_{- 0.27}$ &      $52.77^{+ 0.11}_{- 0.30}$ &      $ 0.35^{+ 0.22}_{- 0.17}$ &                   $= {\rm Fe}$ &                   $= {\rm Fe}$ &     $     512.3 /       482$ \\
 180 &      M1+M2 &      $ 2.38^{+ 0.62}_{- 0.87}$ &      $ 6.70^{+ 0.09}_{- 0.06}$ &      $ 7.39^{+ 0.35}_{- 0.23}$ &              $\cdot\cdot\cdot$ &      $54.09^{+ 0.26}_{- 0.76}$ &      $53.23^{+ 0.31}_{- 0.36}$ &              $\cdot\cdot\cdot$ &      $ 0.01^{+ 0.14}_{- 0.01}$ &      $ 0.09^{+ 0.82}_{- 0.06}$ &      $ 0.21^{+ 0.34}_{- 0.10}$ &     $     213.4 /       180$ \\
\hline
\end{tabular}
\normalsize
\end{center}
 
All the errors are at 68\% confidence level. \\
$^{\rm a}$~Instruments whose spectra were considered in the best-fit procedure. \\
$^{\rm b}$~Abundances are referred to solar photospheric values of \citet{AndersGrevesse1989}. \\
\vfill
\end{minipage}
\end{sidewaystable*}

We have adopted as best fit model an absorbed optically thin plasma with a minimum number of thermal components suitable to describe reasonably well the observed spectrum. The fit procedure has been performed by linking together the abundances of elements heavier than He and leaving as free parameters the temperatures, the emission measures, the hydrogen column density, and the metallicity. After this initial step, we have also left as free parameters the abundances of individual elements which significantly improved the fit. The errors have been estimated by checking $\Delta\chi^{2}$ while varying simultaneously all the best fit parameters. In Table~\ref{tab:fitres} we list parameters and errors of the best fit models and in Fig.~\ref{fig:epicspec} we display observed and predicted X-ray spectra.

The hydrogen column densities derived from X-ray spectral analysis range from 0.32 to $2.4\times10^{21}\,{\rm cm^{-2}}$. The average value, $\sim1.5\times10^{21}\,{\rm cm^{-2}}$, has been assumed for the spectra with low signal to noise ratio. This $N_{\rm H}$ range is compatible with the values derived by \citet{SciortinoDamiani1998} from the analysis of {\it ROSAT}/PSPC spectra, and is compatible with the $A_{V}$ value of $0.1-0.8$ estimated by \citet{WalterVrba1994}. The derived $N_{\rm H}$ does not vary to much from star to star suggesting that only small contribution to the hydrogen column density is due to local circumstellar material.

\subsection{X-ray flux and luminosity}
\label{xrayflux}

From the best fit models of the X-ray brightest USco members we have evaluated their unabsorbed X-ray flux and luminosity. These flux values have been used to derive a mean conversion factor from PN equivalent count rate to flux of $2.37\times10^{-12}\,{\rm erg\,cm^{-2}}$ in the $0.5-8.0$\,keV energy band. By this factor we have derived X-ray fluxes also for the faintest X-ray USco sources. This procedure relies on the hypothesis that plasma characteristics do not vary substantially from source to source. Since source~100 displays much harder spectra than other USco sources (see Sect.~\ref{xraycol}), we have not considered it for the conversion factor evaluation. X-ray fluxes and luminosities of the photometric USco members are listed in Table~\ref{tab:memb}.


\section{Variability analysis}
\label{uscovar}

\subsection{Kolmogorov-Smirnov test}

We have studied the X-ray variability of all the detected sources by applying the unbinned Kolmogorov-Smirnov (KS) test. We have analyzed the photon arrival times adding the PN, MOS1 and MOS2 data, because we aimed at maximizing the $S/N$ ratio of each source and therefore at increasing the diagnostic power for the weakest sources. The joint use of different instruments, reasonable because they share the same energy band, has been performed by intersecting the good time intervals of the three instruments. The procedure has the advantage of increasing the total counts of each source by a factor $\sim2$, losing only 10\% of the time coverage, mainly due to the lower PN exposure\footnote{The start of the PN detector is usually delayed of few ks ($\sim2.5$ in our cases) with respect to the two MOS detectors.}.

Since the KS test probes unbinned arrival times, it does not distinguish between source and background photons. Hence to check background variability we have also applied the KS test to background events of each source.

\begin{table}[b]
\begin{center}
\caption{Number of variable sources derived from the Kolmogorov-Smirnov test.}
\label{tab:ksres}
\begin{tabular}{ccccc}
\hline\hline
                               & \multicolumn{2}{c}{All Sources}           &  \multicolumn{2}{c}{USco Members}          \\
\multicolumn{1}{c}{Conf. Lev.} & $N_{\rm var}$ & $N_{\rm var}/N_{\rm tot}$ & $N_{\rm var}$ & $N_{\rm var}/N_{\rm tot}$ \\
\hline
90\%                           & 42            & 18.8\%                    & 14            & 63.6\%      \\
99\%                           & 19            & 8.5\%                     & 13            & 59.1\%      \\
\hline
\end{tabular}
\end{center}
\end{table}

We have adopted the confidence levels of 90\% and 99\% for testing the hypothesis of constant emission. Table~\ref{tab:ksres} reports the results obtained for the sample of all detected X-ray sources, and for the selected subset of photometric USco members. Among the sources which turned out to be variable at 99\% we have discarded source~97 (not an USco member) because its variability is due to contamination: source~97 is located near the strong source~100, characterized by an intense flare (see Sect.~\ref{flranalysis}), whose large point spread function affects the photons extracted in the source~97 region. Some of the sources variable at 90\% confidence level have local background variable at the same confidence level, however we have not discarded them because the cumulative functions of source and background photon arrival times do not appear to be similar.

\begin{figure*}
\centering
\includegraphics[width=17cm,height=22cm]{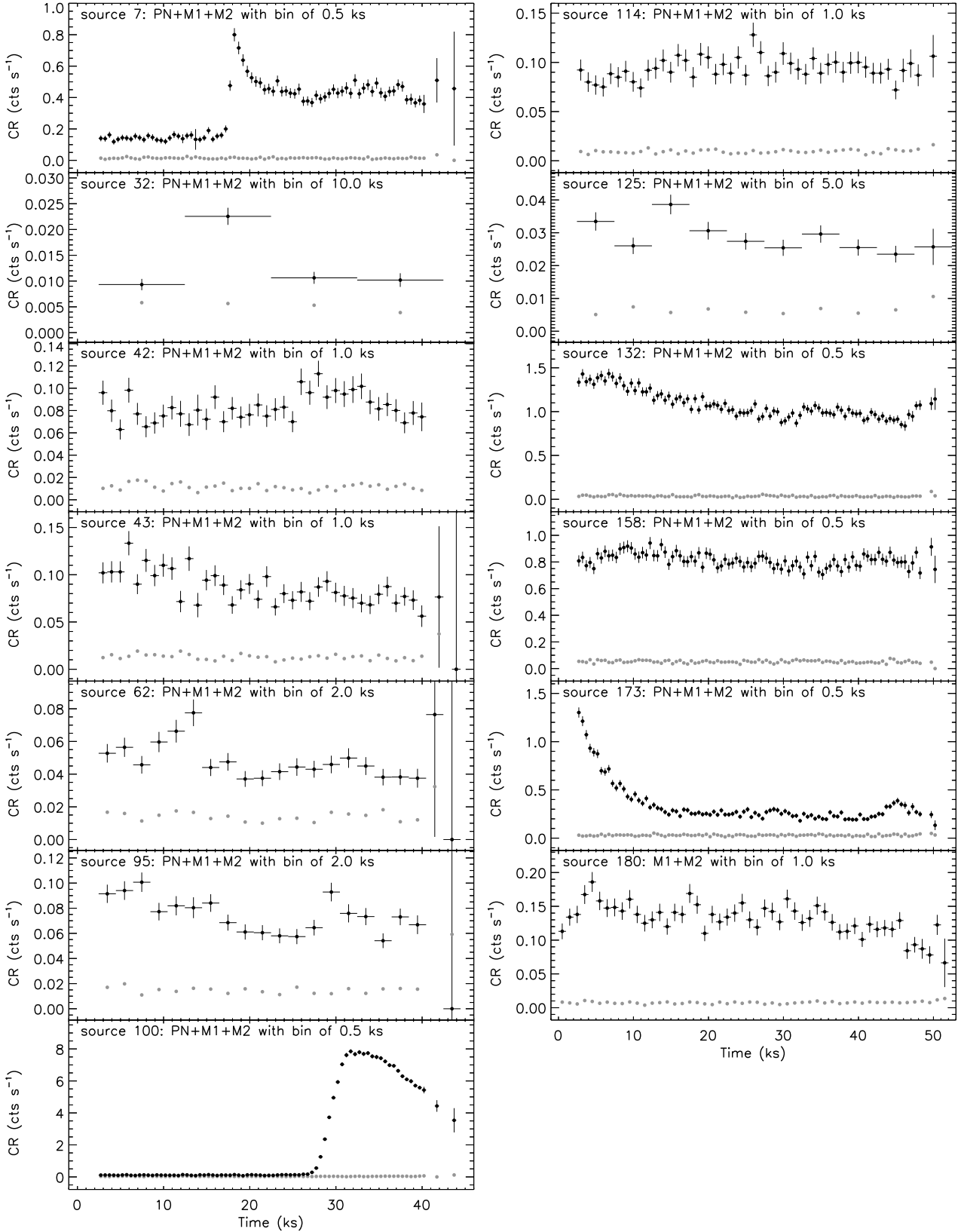}
\caption{Light curves of the USco members which turned out to be variable at the 99\% confidence level by applying the KS test. Light curves are obtained by adding the counts of all the EPIC instruments. Light curves are not background subtracted, background level is shown by gray dots, and it is constant in all cases. Arrival times are referred to the observations start. Sources displayed on the left column are those of field~2, right column contains sources of field~1.}
\label{fig:epiclc}
\end{figure*}

As indicated in Table~\ref{tab:ksres} we have found 13 USco stars variable at 99\% (i.e. 59.1\% of the detected USco members). In Fig.~\ref{fig:epiclc} we present the light curves of these 13 USco members. We have plotted light curves without background subtraction; the background level for each source is shown by gray dots.

These 13 X-ray variable USco sources are also the 13 X-ray brightest among the 22 USco members. The weakest of this subsample is source~32 which has 290 EPIC net counts. Therefore our results are likely influenced by statistics and the estimated fraction of USco variable sources represents a lower limit.

The X-ray emission of the USco brown dwarf (source~45, 247 total EPIC counts, $\sim140$ of which are background) does not display significant variability, suggesting that the observed X-rays are probably not due to isolated and bright flaring events.

\subsection{Flare analysis}
\label{flranalysis}

Figure~\ref{fig:epiclc} shows that the four sources 7, 32, 100, and 173 display intense and isolated flare events. The most intense flare was detected from source 100, whose count rate increased by a factor 100 during the flare. Source 7 shows a peculiar behavior: after the flare maximum the source reaches a quiescent emission higher by a factor $\sim2$ than before the flare. A similar light curve has been reported by \citet{TsuboiKoyama1998} who observed the X-ray emission from the WTTS V773, indicating hence that this kind of flare may be common on non-accreting PMS stars.

The analysis of flares allows to infer the length of the flaring structure. We have adopted the method proposed by \citet{RealeBetta1997} and recently used by \citet{FavataFlaccomio2005}. In this section we apply this method to the flares of sources 7, 100, and 173, in order to infer characteristic lengths and heating properties of these flaring coronal structures.

Source 132 shows, during the entire observation, a slow decline (see Fig.~\ref{fig:epiclc}). Different phenomena may produce this light curve: the superposition of small amplitude variability, rotational modulation, or the decay of a flare. We have made this latter hypothesis, assuming therefore that the observation start coincided with the flare maximum and the subsequent decline was the flare decay. Hence we have performed, also for this source, time resolved spectral analysis to derive the length of the flaring loop. We have no means of proving the correctness of this assumption, and therefore all the derived results depend on the validity of this hypothesis.

The \citeauthor{RealeBetta1997} method relies on the dependence of flaring loop cooling on its length: the longer the decay, the longer the loop. The flare decay is also regulated by the amount of energy released into the loop itself after the flare maximum. This amount can be probed by studying the evolution of the plasma temperature and density during the flare decay.

The assumptions of the \citeauthor{RealeBetta1997} method are: the flaring structures is a semicircular and symmetric coronal loop with uniform cross section whose geometry does not change significantly during the flare; the plasma is confined inside the flaring loop; the plasma motion and the energy transport can occur only along the magnetic field. Under these assumptions it is possible to derive the loop length from the relation:

\begin{equation}
L_{9}=\frac{\tau_{LC} \sqrt{T_{7}}}{120 f(\zeta)}
\label{eq:looplength}
\end{equation}

\noindent
where: $L_{9}$ is the loop semi-length in $10^{9}\,{\rm cm}$, $\tau_{LC}$ is the decay time, in seconds, of the light curve; $T_{7}$ is the peak temperature reached during the flare, in units of $10^{7}$\,K; $\zeta$ is the slope of the trajectory followed by the flaring plasma in the $\log T$ vs. $\log N_{\rm e}$ space, during the decay. Under the hypothesis of constant volume of the flaring structures the electron density $N_{\rm e}$ is proportional to the square root of the emission measure {\it EM}. The function $f(\zeta)$ takes into account the amount of heating released into the loop during the decaying phases, and it is defined as the ratio between the spontaneous thermodynamic loop decay $\tau_{th}$ and that measured from the light curve $\tau_{LC}$. Note that $f(\zeta)$ depends on the instrument used to observe the flare X-ray emission. The calibrated $f(\zeta)$ function for the EPIC/PN detector, provided by \citet{RealeGudel2004}, is:

\begin{equation}
f(\zeta)=\frac{\tau_{LC}}{\tau_{th}}=a_{1}\exp(-\zeta/a_{2})+a_{3}
\label{eq:fzeta}
\end{equation}
\noindent
with $a_{1}=11.6$, $a_{2}=0.56$, and $a_{3}=1.2$.
 
\subsubsection{Fitting procedure}

All the physical quantities contained in Eq.~(\ref{eq:looplength}) refer to the loop where the flare occurred. To separate the X-ray emission of the quiescent corona from that of the flaring structure, we have estimated a model which describes the quiescent phase preceding, or succeeding, the flare, assuming that this quiescent component has not changed during the flare (for the selection of the source~132 quiescent phase see Sect.~\ref{source132flare}). Hence we have fitted the EPIC spectra of the flaring phases adding to the model of the quiescent corona another thermal component, which takes into account only the flaring loop.

We have fitted the quiescent phase with an absorbed optically thin plasma with 1 or 2 thermal components, with all the abundances of elements heavier than He tied together.

For all the flaring phases we have added to the above quiescent model a further absorbed optically thin isothermal plasma with all the abundances of elements heavier than He tied to Fe. In the best fit procedure all the parameters of the quiescent model have been kept frozen. We have instead left as free parameters the temperature, emission measure, and metallicity of the added flaring model. In all cases we have found that the hydrogen column density $N_{\rm H}$ remains constant and compatible with the value estimated from the analysis of the whole spectrum (see Table~\ref{tab:fitres}), hence it was kept frozen. The error estimation was performed by studying the $\Delta\chi^{2}$ while changing simultaneously all the best fit parameters.

\begin{figure}
\resizebox{\hsize}{!}{\includegraphics{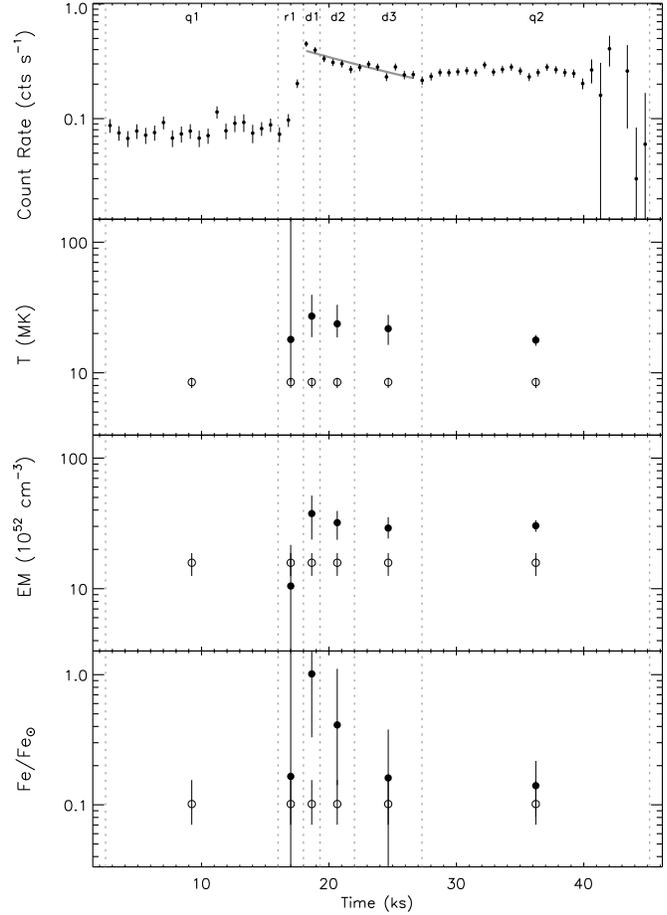}}
\caption{Flare analysis of source~7. {\it Upper panel}: background subtracted light curve obtained by binning PN arrival times with bin of 700\,s; the solid gray line represents the best fit exponential decay. {\it Lower panels}: plasma temperature, emission measure, and metallicity \protect{\citep[referred to][]{AndersGrevesse1989}} derived from spectral fitting of each time interval; open symbols mark the quiescent component, filled symbols the flaring component.}
\label{fig:fitres7}
\end{figure}

\begin{figure}
\resizebox{\hsize}{!}{\includegraphics{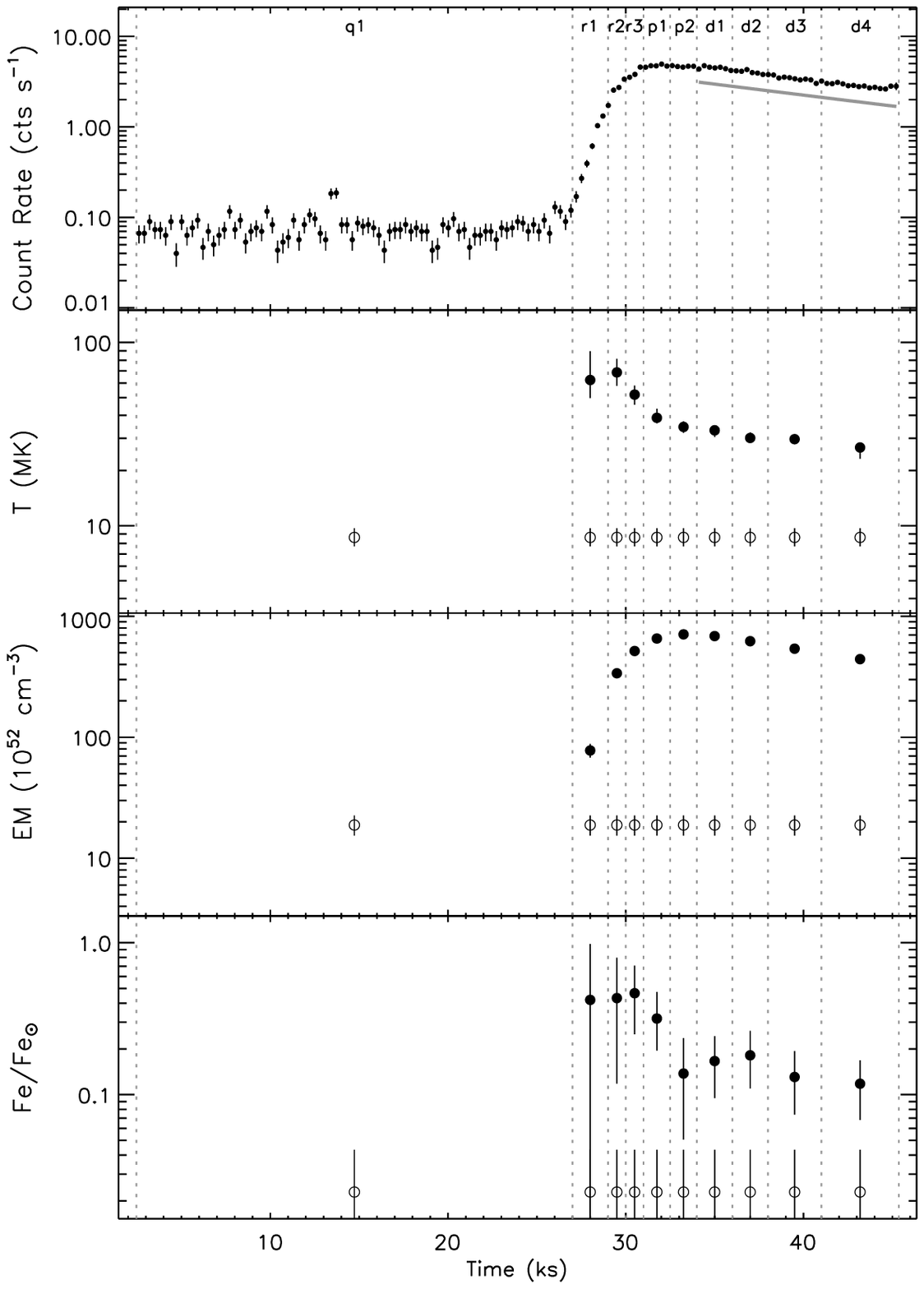}}
\caption{Flare analysis of source~100. {\it Upper panel}: background subtracted light curve obtained by binning PN arrival times with bin of 300\,s; the solid gray line represents the exponential decay (reduced by a factor 1.5 to make it visible) obtained from best fit of count rates in the d1, d2, d3, and d4 intervals. {\it Lower panels}: plasma temperature, emission measure, and metallicity derived from spectral fitting of each time interval; open symbols mark the quiescent component, filled symbols the flaring component.}
\label{fig:fitres100}
\end{figure}

\subsubsection{Source~7}

Source 7 is a M1.5 WTTS, with $M_{\star}\approx0.3\,M_{\odot}$ and $R_{\star}\approx2.3\,R_{\odot}$ \citep{AdamsWalter1998}. Its X-ray light curve is plotted in the upper panel of Fig.~\ref{fig:fitres7}. We have divided the whole observation into six intervals. The quiescent spectrum, is collected during the ``q1'' interval. The results of the best fit are displayed in the lower panels of Fig.~\ref{fig:fitres7} and listed in Table~\ref{tab:flare7}. Significant abundance variations have been registered: the flaring plasma reaches an abundance similar to the solar photospheric value, while the quiescent coronal plasma is heavily metal depleted.

\begin{figure}
\resizebox{\hsize}{!}{\includegraphics{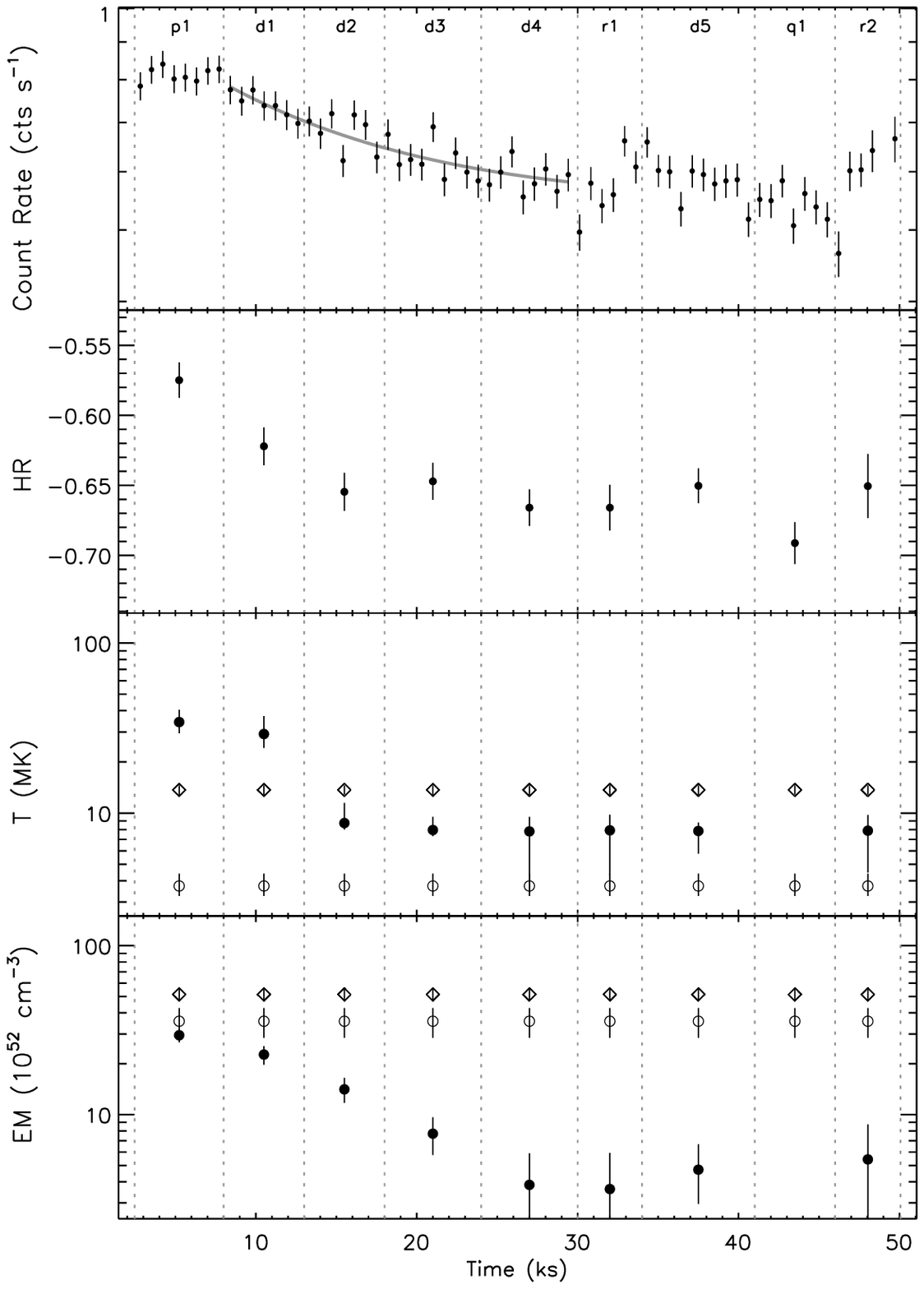}}
\caption{Flare analysis of source~132. {\it From upper to lower panel}: background subtracted PN  light curve with bin of 700\,s, solid gray line represents the exponential decay obtained from best fit of count rates in the d1, d2, d3, and d4 intervals; hardness ratio $HR=(H-S)/(H+S)$, with $S$ and $H$ indicating the number of photons detected in the $0.3-1.5$ and $1.5-7.9$\,keV band, respectively; plasma temperature, emission measure, and metallicity derived from spectral fitting of each time interval;  open symbols mark the quiescent component, filled symbols the flaring component.}
\label{fig:fitres132}
\end{figure}

\begin{figure}
\resizebox{\hsize}{!}{\includegraphics{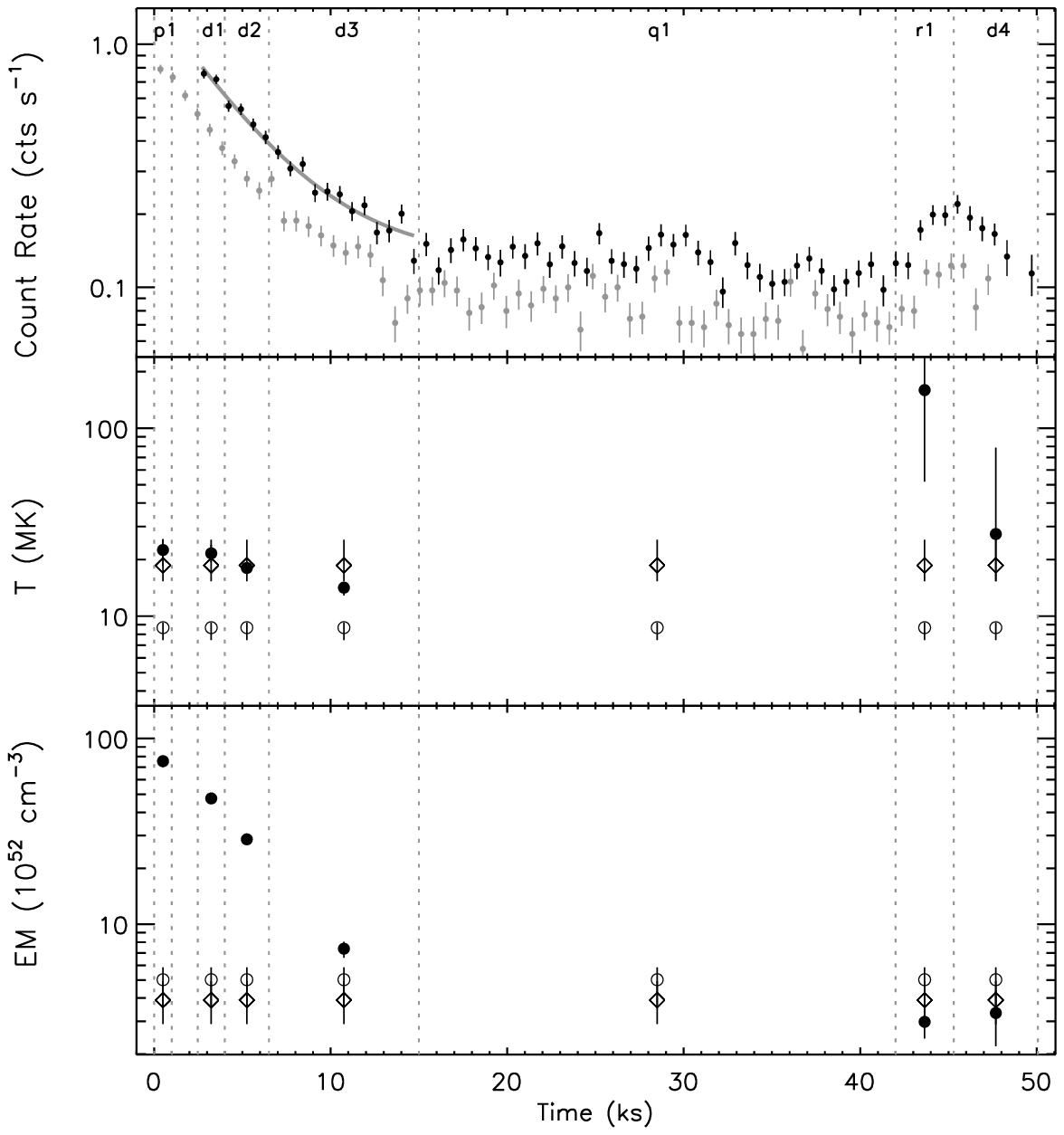}}
\caption{Flare analysis of source~173. {\it Upper panel}: background subtracted PN light curve (black) and MOS light curve (gray) obtained with bin of 700\,s; the  solid gray line represents the best fit exponential decay. {\it Lower panels}: plasma temperature and emission measure derived from spectral fitting of each time interval; open symbols mark the quiescent component, filled symbols the flaring component. All the best fit results refer only to PN data analysis, except for fit results of the p1 interval that were obtained from MOS data (more details can be found in Sect.~\ref{flranalysis}).}
\label{fig:fitres173}
\end{figure}

The decay time of the flare has been derived by fitting the count rates in the d1, d2, and d3 intervals, assuming as best fit model an exponential function plus a constant. The constant ($0.08\,{\rm cts\,s^{-1}}$), representing the quiescent X-ray emission, has been obtained as the average count rate value of the quiescent phase q1. The derived exponential decay of the best fit function (shown in the upper panel of Fig.~\ref{fig:fitres7}) is $\tau_{LC}=11.1\pm1.3$\,ks.

\subsubsection{Source~100}
\label{source100flare}

Counterpart of source~100 is HD~142578, whose optical photometry \citep{SlawsonHill1992} suggests $M_{\star}\sim2M_{\odot}$ and $R_{\star}\sim3\,R_{\odot}$. The {\it XMM-Newton} observation of field~2, which contains source 100, is affected by high background levels especially in the last $\sim5$\,ks of the exposure. Hence the time screening procedure made us discard mainly the last section of the observation, as can be seen from the light curves contained in the left column of Fig.~\ref{fig:epiclc}. The flare of source~100 occurred $\sim30$\,ks after the observation start, and only the first decay phases were recorded. To optimize the analysis of the flare decay, we have considered the whole observation exposure without applying any time screening. The high background level registered in the last part of the field~2 observation does not heavily affect the analysis of source 100 because the source count rate is much higher than background count rate.

In the upper panel of Fig.~\ref{fig:fitres100} we display the background-subtracted PN light curve of source 100 extracted from the unscreened field~2 observation. We have divided the observation into 10 intervals. Both the quiet and flaring emission are well described by an absorbed 1-$T$ thermal component. The results on the evolution of temperature, emission measure, and global abundance are reported in the lower panels of Fig.~\ref{fig:fitres100} and in Table~\ref{tab:flare100}. Also in this case we have registered a significant abundance increase: the flaring plasma shows a metallicity higher than the quiescent one by a factor $\sim20$.

A flare decay time of $15.4\pm0.4$\,ks has been obtained by fitting the ${\rm d1 - d4}$ intervals. Best fit model (reduced by a factor 1.5 to make it visible) is shown with solid gray line in Fig.~\ref{fig:fitres100}.

\begin{figure*}
\resizebox{0.5\hsize}{!}{\includegraphics{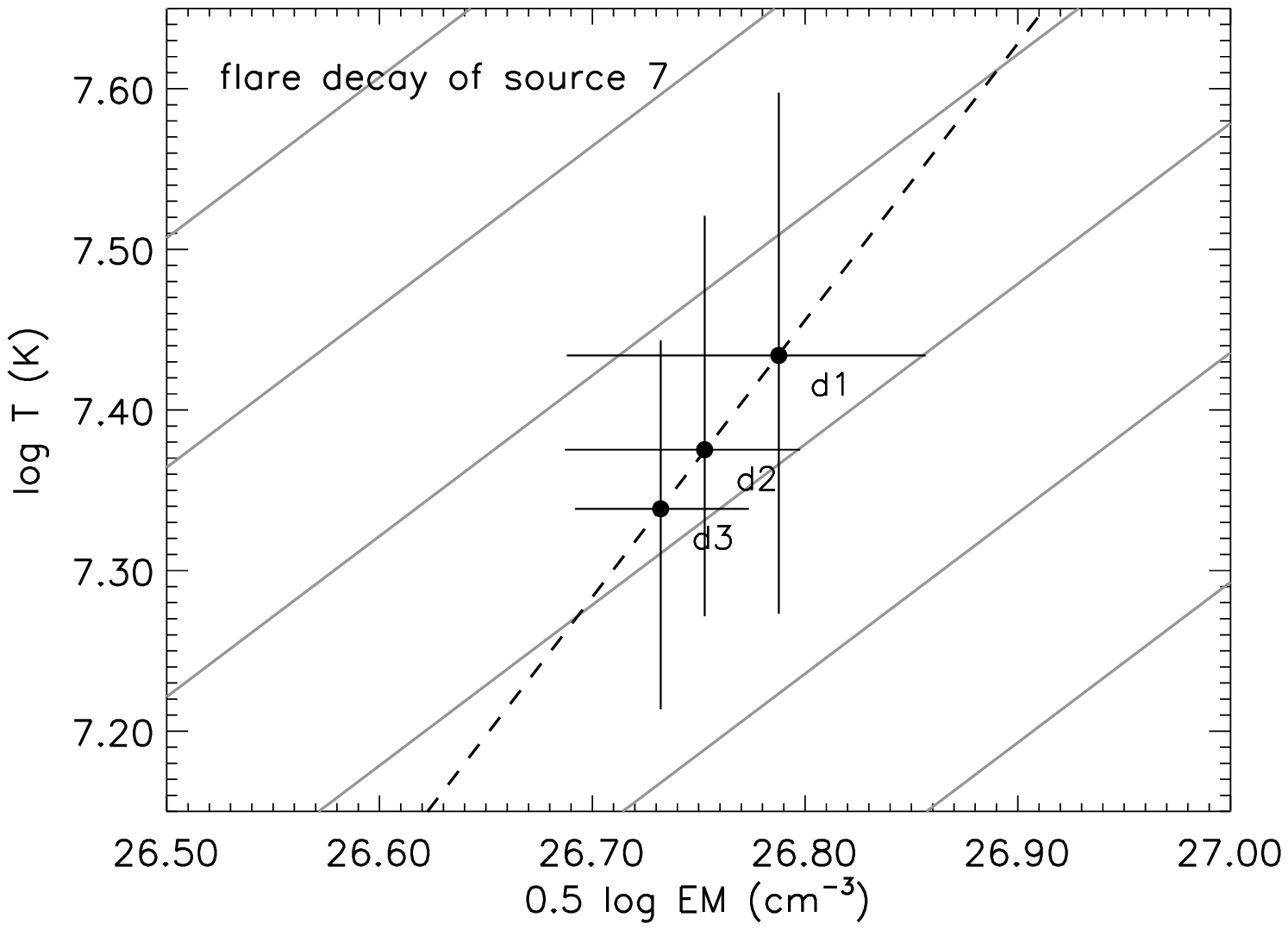}}
\resizebox{0.5\hsize}{!}{\includegraphics{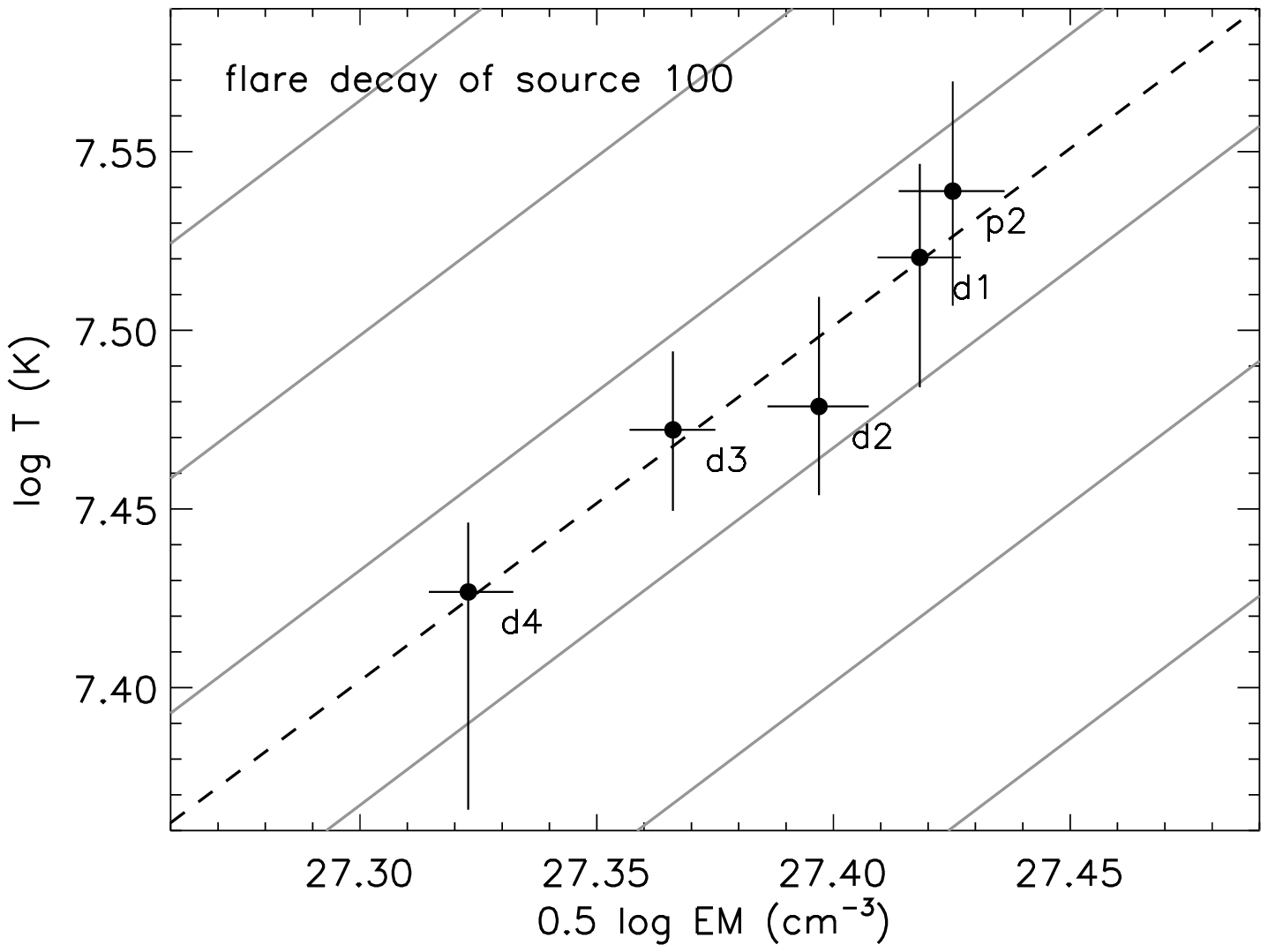}}

\resizebox{0.5\hsize}{!}{\includegraphics{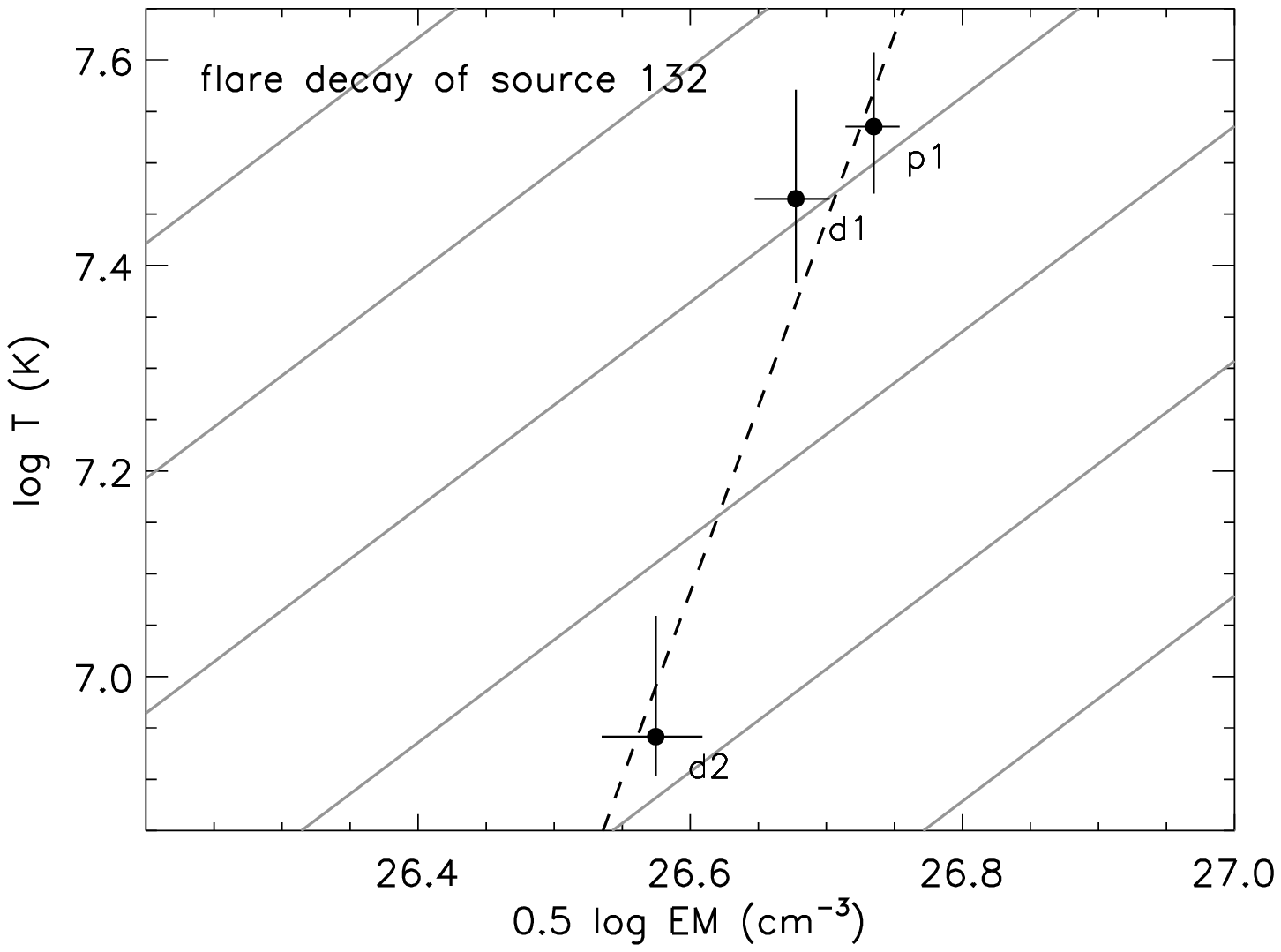}}
\resizebox{0.5\hsize}{!}{\includegraphics{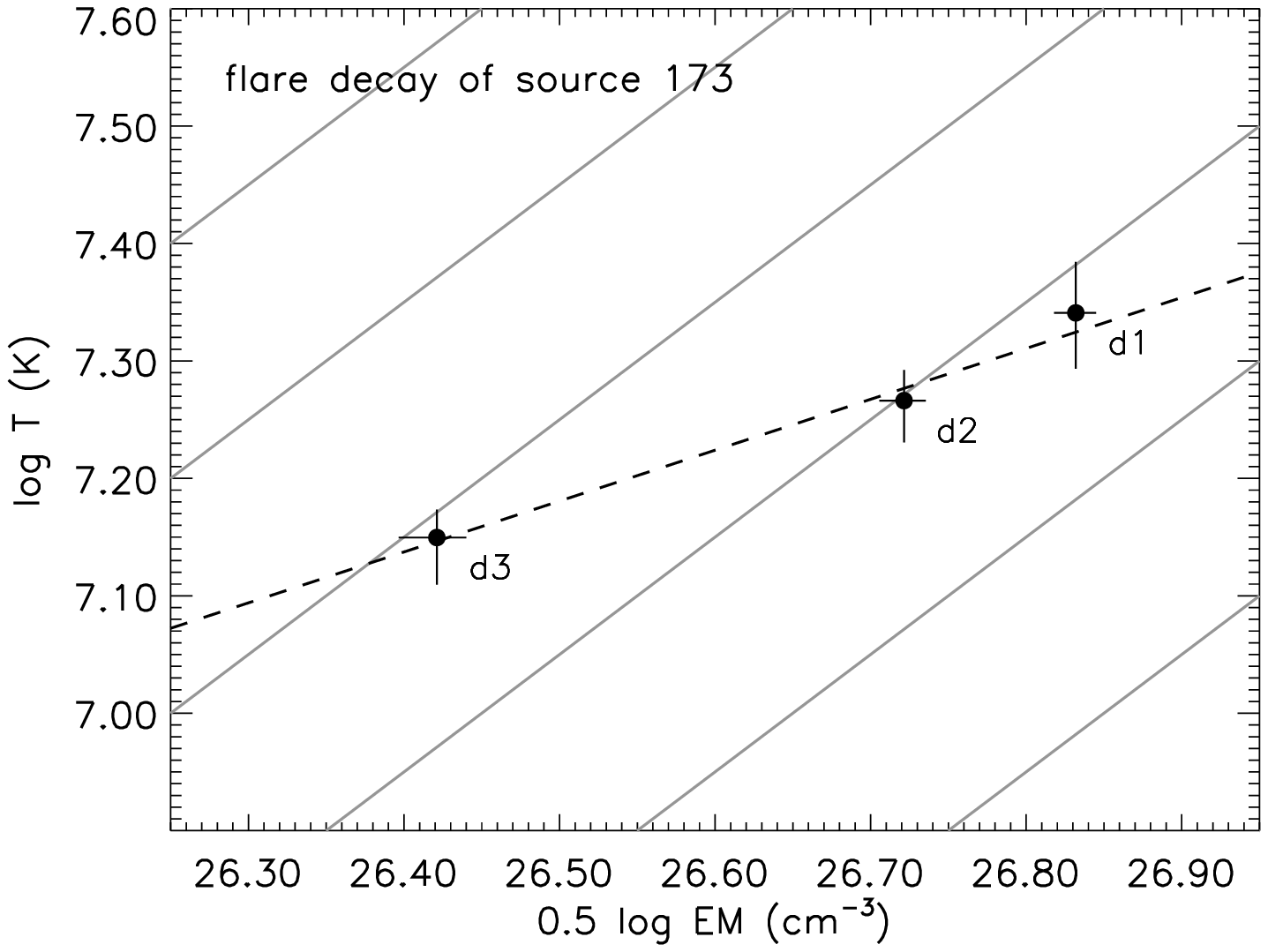}}

\caption{Evolution of the flare decaying phases in the temperature vs. density diagram ($0.5\log {\it EM} \propto \log N_{\rm e}$, where $N_{\rm e}$ is the electron density). Dotted black lines mark the linear best fit model, solid gray lines indicate lines with slope 1.}
\label{fig:tempvsdens}
\end{figure*}

\subsubsection{Source~132}
\label{source132flare}

Source~132 is a K0 young star, whose $N$-band excess and asymmetric H$\alpha$ features suggest to be actively accreting material from its circumstellar disk \citep{MamajekMeye2004}. From the $L_{\rm bol}$ and $T_{\rm eff}$ values derived by \citet{PreibischZinnecker1999} we deduce $R_{\star}\sim3.5\,R_{\odot}$. The upper panel of Fig.~\ref{fig:fitres132} shows the background-subtracted PN light curve of source~132. The decay phase lasts for the first $\sim 30$\,ks, after that variability with smaller amplitude affects the emission.

By inspection of the light curve we have divided the whole observation into 9 segments, as indicated in Fig.~\ref{fig:fitres132}. To individuate the interval to adopt for the quiescent emission we have evaluated the hardness ratio $HR=(H-S)/(H+S)$ of each interval, where $S$ indicates the number of photons detected in the $0.3-1.5$\,keV band, and $H$ those in the $1.5-7.9$\,keV band. The results are displayed in the second panel of Fig.~\ref{fig:fitres132}. We have chosen the interval with the softest spectrum for the estimation of quiescent model, noting that this interval, named q1, also shows the minimum count rate.

We have checked that the abundance of the flaring plasma does not vary and was compatible with that of the quiescent model. We have therefore kept it fixed to the value derived from the analysis of the whole spectrum: ${\rm Fe = 0.21\,Fe_{\odot}}$. In the lower panels of Fig.~\ref{fig:fitres132} we report the best fit values of temperature and emission measure for each explored time interval. All the best fit parameters are also listed in Table~\ref{tab:flare132}. The hypothesis of a flare decay to describe the observed light curve is supported by the decrease of the observed temperature. The decay time ($11.4\pm1.3$\,ks) of the light curve has been estimated by fitting the observed count rates of the ${\rm d1-d4}$ intervals.

\subsubsection{Source~173}

\begin{table*}
\begin{center}
\caption{Results of loop length analysis.}
\label{tab:looplength}
\begin{tabular}{lcccc}
\hline\hline
Usco Star                             & 7               & 100             & 132               & 173 \\
\hline 
\multicolumn{5}{c}{Stellar Properties}        \\
\hline 
$V$                                   & 13.01$^{\rm a}$ & 8.63$^{\rm b}$  & $\cdots$          & $\cdots$         \\
$B-V$                                 & 1.47$^{\rm a}$  & 0.23$^{\rm b}$  & $\cdots$          & $\cdots$         \\
$R_{\star}\,(R_{\odot})$              & 2.3$^{\rm c}$   & $\sim3^{\rm d}$ & $\sim3.5^{\rm e}$ & $\sim1^{\rm f}$  \\
\hline 
\multicolumn{5}{c}{Flaring Loop Properties}   \\
\hline 
$T_{\rm obs}$\,(MK)                   & $27\pm10$       & $69\pm12$       & $34\pm 5$         & $23\pm 3$        \\
$T_{\rm peak}$\,(MK)                  & $46 \pm20$      & $132\pm25$      & $60 \pm11$        & $38 \pm 6$       \\
$\zeta$                               & $1.7\pm4.6$     & $1.0\pm0.4$     & $3.6\pm1.2$       & $0.43\pm0.12$    \\
$f(\zeta)$                            & $1.7 \pm4.3$    & $3.2 \pm1.5$    & $1.22\pm0.04$     & $6.5 \pm1.2$     \\
$\tau_{LC}$\,(ks)                     & $11.1\pm1.3$    & $15.4\pm0.4$    & $11.4\pm1.3$      & $3.96\pm0.15$    \\
$L\,{\rm (10^{10}\,cm)}$              & $\sim 11$       & $15\pm7$        & $19\pm3$          & $0.98 \pm 0.19 $ \\
$L\,(R_{\star})$                      & $\sim0.7$       & $\sim0.7$       & $\sim0.8$         & $\sim0.14$       \\
$N_{\rm e}\,{\rm (10^{10}\,cm^{-3})}$ & $\sim7.0$       & $\sim21$        & $\sim3.8$         & $\sim400$        \\
$B$\,(G)                              & $\sim150$       & $\sim370$       & $\sim110$         & $\sim1000$       \\
\hline
\end{tabular}

\end{center}
The electron density $N_{\rm e}$ of the flaring loop is evaluated as $\sqrt{EM_{\rm max}/(0.8\cdot2\pi\beta^2L^3)}$, assuming $\beta=0.1$. The magnetic field values $B$ represent the minimum values needed to confine the flaring plasma at the maximum pressure, and are evaluated as $\sqrt{8\pi P_{\rm max}}$, with $P_{\rm max}=k2(N_{\rm e}T)_{\rm max}$.

$^{\rm a}$~From \citet{WalterVrba1994}. 

$^{\rm b}$~From \citet{SlawsonHill1992}. 

$^{\rm c}$~From \citet{AdamsWalter1998}. 

$^{\rm d}$~Estimated from the $B-V$ color and the \citet{SiessDufour2000} models.

$^{\rm e}$~Estimated from the $L_{\rm bol}$ and $T_{\rm eff}$ values derived by \citet{PreibischZinnecker1999}.

$^{\rm f}$~Estimated from the $H$-$K$ 2MASS color and the \citet{SiessDufour2000} models.

\end{table*}

No bibliographic information is available for the 2MASS and DENIS counterparts of source~173, which we have selected as photometric USco member for the first time. The 2MASS $H$-$K$ color and the \citet{SiessDufour2000} models indicate $R_{\star}\sim1\,R_{\odot}$. The background-subtracted PN light curve of source~173 is shown in black in the upper panel of Fig.~\ref{fig:fitres173}. However the PN observation started when the flare was already in the decay phase. To determine the length of the flaring structure we need to measure the maximum temperature of the plasma during the flare (see Eq.~(\ref{eq:looplength})). To this aim we have inspected the MOS1 and MOS2 data because their observation started $\sim2.5$\,ks before that of PN detector. In the upper panel of Fig.~\ref{fig:fitres173} we also report the background-subtracted light curve obtained by adding MOS1 and MOS2 photon arrival times. From the comparison of PN and MOS light curves it emerges that the PN detector missed a significant fraction of the flare evolution. We assume that the MOS data started just after the flare maximum. However, since the maximum temperature is usually reached during the rising phases (see the temperature evolution of source~100 shown in Fig~\ref{fig:fitres100}) we are likely determining a lower limit.

We have divided the PN observation into 6 intervals, as reported in Fig.~\ref{fig:fitres173}, with another interval which probe the flare peak from the MOS data with duration of 1\,ks. The quiescent emission has been estimated from the constant level reached after the flare end (q1 interval).

Global metal abundance remains constant during the whole observation. Hence we have performed the spectral fitting leaving it fixed to the values of ${\rm Fe/Fe_{\odot}}=0.35$ (see Table~\ref{tab:fitres}). The best results are shown in the lower panels of Fig.~\ref{fig:fitres173} and listed in Table~\ref{tab:flare173}. The best fit function of the flare decay (d1, d2, and d3 intervals) is displayed with a solid gray line in the upper panel of Fig.~\ref{fig:fitres173}, it has a decay time $\tau_{LC}$ of $3.96\pm0.15$\,ks.

\subsubsection{Loop length estimation}

To estimate the size of the flaring loops we need to determine the slope of the path described in the density vs. temperature plane by the plasma during the cooling phases. Under the hypothesis of constant volume of the loop, the emission measure is proportional to the square of the electron density.

In Fig.~\ref{fig:tempvsdens} we show the evolution of $\log T$ and $0.5 \log {\it EM}$ for the flares analyzed. For each flare we report only the points regarding the decay phases. In all the cases the hypothesis of a linear trajectory is verified. We have performed linear regression of the plotted points, and the best fit functions are displayed by black dashed lines. We have also marked with solid gray lines loci characterized by slope 1, in order to easily compare the slopes in the plots. The estimated slopes $\zeta$ are $1.7\pm4.6$, $1.0\pm0.4$, $3.6\pm1.2$, and $0.43\pm0.12$ for the source 7, 100, 132, and 173, respectively. Then we have evaluated $f(\zeta)$ by applying Eq.~(\ref{eq:fzeta}). In Table~\ref{tab:looplength} we report the measured $\zeta$ and the corresponding $f(\zeta)$. The large $f(\zeta)$ values obtained for source 100 and 173 suggest the presence of sustained heating. 

To evaluate the loop maximum temperature experienced during the flare we have converted the measured maximum temperature adopting the relation:
\[
T_{\rm peak} = 0.184\,T_{\rm obs}^{1.130}
\label{eq:tpeak}
\]
where $T_{\rm peak}$ and $T_{\rm obs}$ are expressed in K \citep{RealeGudel2004}. Both $T_{\rm obs}$ and $T_{\rm peak}$ are listed in Table~\ref{tab:looplength}.

The estimated loop semi-lengths, for sources 7, 100, 132, and 173, are $\sim0.7$, $\sim0.7$, $\sim0.8$, and $\sim0.14$ of stellar radii (see Table~\ref{tab:looplength}).

Starting from the loop size we have derived indication on the plasma density and magnetic field. We define $\beta$ as the ratio between the section radius $r$ and the loop semi-length $L$. The loop volume is therefore:
\begin{equation}
V_{\rm loop}=2\pi\beta^2L^3
\label{eq:loopvolume}
\end{equation}
Considering the highest value of {\it EM}, and the loop volume given by (\ref{eq:loopvolume}), we can evaluate the peak density reached by the plasma:
\begin{equation}
N_{\rm e}=\sqrt{\frac{{\it EM}_{\rm max}}{0.8\cdot2\pi\beta^2L^3}}
\label{eq:loopdens}
\end{equation}
where $N_{\rm e}$ is the electron density, and the factor 0.8 has been assumed as the ratio between $N_{\rm H}$ and $N_{\rm e}$. From density and temperature we can estimate the pressure. Since the confinement of the plasma into the loop is due to the magnetic field we can infer the minimum magnetic field needed to constrain the plasma, when it reaches the maximum pressure during the flare. The magnetic field $B$ in gauss is given by:
\begin{equation}
B=\sqrt{8\pi k2(N_{\rm e}T)_{\rm max}}
\label{eq:loopb}
\end{equation}
Solar coronal loops usually display $\beta\approx0.1$, while in some active stars larger values ($\beta\approx0.3$) have been inferred \citep{FavataMicela2000}. Assuming $\beta=0.1$ we have derived the values listed in Table~\ref{tab:looplength}. The density $N_{\rm e}$ and the magnetic field $B$ are evaluated considering the highest values of {\it EM} and $P$, respectively, which usually did not occur in the same time interval. Magnetic field values larger than a few $10^{2}$\,G are compatible with typical photospheric values of young late type stars \citep{GuentherLehmann1999,SymingtonHarries2005}.

\subsection{Long term variability: comparison with ROSAT results}

The field~2 was previously observed with a pointed {\it ROSAT}/PSPC observation in 1993 \citep{SciortinoDamiani1998}. Hence the comparison between {\it XMM-Newton} and {\it ROSAT} data allows to search long term X-ray variability of USco members.

\citet{SciortinoDamiani1998} evaluated X-ray luminosity in the $0.2-2.0$\,keV energy band assuming a distance of 160\,pc, hence we have scaled the {\it ROSAT}/PSPC luminosity to a distance of 145\,pc. We have evaluated $L_{\rm X}$ from {\it XMM-Newton} data in the $0.2-2.0$\,keV energy band by converting the measured X-ray count rate to an unabsorbed energy flux by the conversion factor $2.71\times10^{-12}\,{\rm erg\,cm^{-2}\,cts^{-1}}$ (the procedure adopted to estimate this conversion factor is explained in Sect.~\ref{xrayflux}).

To check for long term variability also for USco members of field~1 we have searched those included in the {\it ROSAT} All Sky Survey Bright Source Catalog \citep[RASS-BSC,][]{VogesAschenbach1999}, performed in 1990-1991. We have found RASS-BSC counterparts for source 114 and 132. Starting from the RASS-BSC count rate and $HR1$, we have evaluated their X-ray fluxes in the $0.1-2.4$\,keV adopting the method of \citet{FlemingMolendi1995}. Then we have applied a conversion factor of 0.80, evaluated with PIMMS V3.6c\footnote{The PIMMS count rate simulator is accessible via URL at {\tt http://heasarc.gsfc.nasa.gov/Tools/w3pimms.html}}, to convert this flux from the $0.1-2.4$\,keV into the $0.2-2.0$\,keV band.

\begin{figure}
\resizebox{\hsize}{!}{\includegraphics{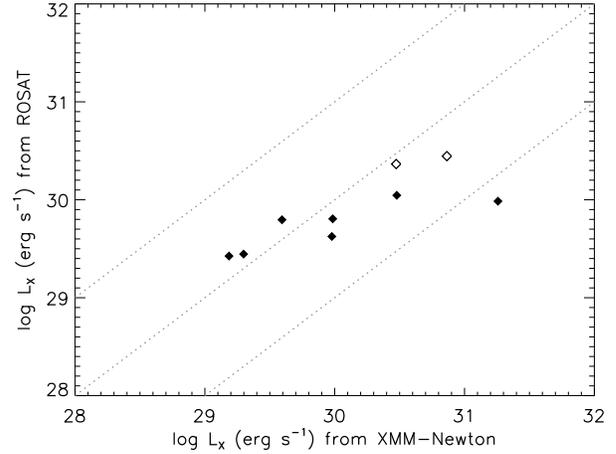}}
\caption{Comparison between the X-ray luminosity of USco members, measured in the $0.2-2.0$\,keV band, measured by {\it ROSAT}/PSPC and {\it XMM-Newton}. Filled diamonds mark USco stars whose {\it ROSAT} X-ray luminosity was estimated by \protect{\citet{SciortinoDamiani1998}} from a pointed observation. Open diamonds indicate USco members whose {\it ROSAT} X-ray luminosity was estimated from the RASS data.}
\label{fig:XMMvsROSATlx}
\end{figure}

The comparison between {\it ROSAT} and {\it XMM-Newton} X-ray luminosity is shown in Fig.~\ref{fig:XMMvsROSATlx}. Filled diamonds indicate source with {\it ROSAT} $L_{\rm X}$ derived from the pointed observation, open diamonds mark sources whose {\it ROSAT} X-ray luminosity was derived from the RASS-BSC.

All the sources, except for source 100, present X-ray luminosity variations smaller than a factor 3; the X-ray luminosity of source 100 measured by {\it ROSAT} is much lower than that we measure in the {\it XMM-Newton} observation, but this difference is entirely due to the flare, in fact the {\it ROSAT} X-ray luminosity is compatible with the quiescent X-ray luminosity of source~100 ($\log L_{\rm X}=30.0$ in the $0.2-2.0$\,keV band).


\begin{figure*}
\resizebox{0.5\hsize}{!}{\includegraphics{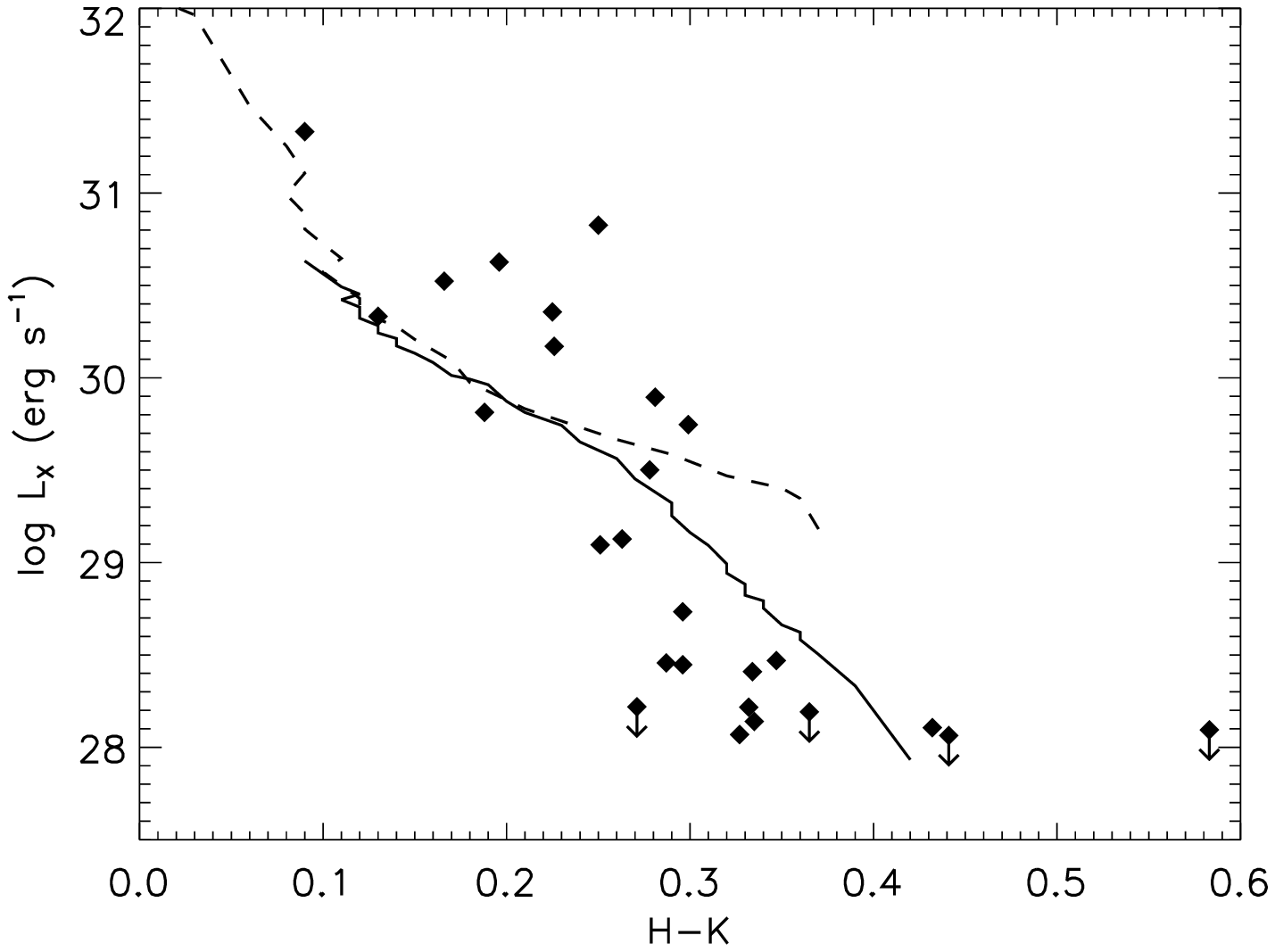}}
\resizebox{0.5\hsize}{!}{\includegraphics{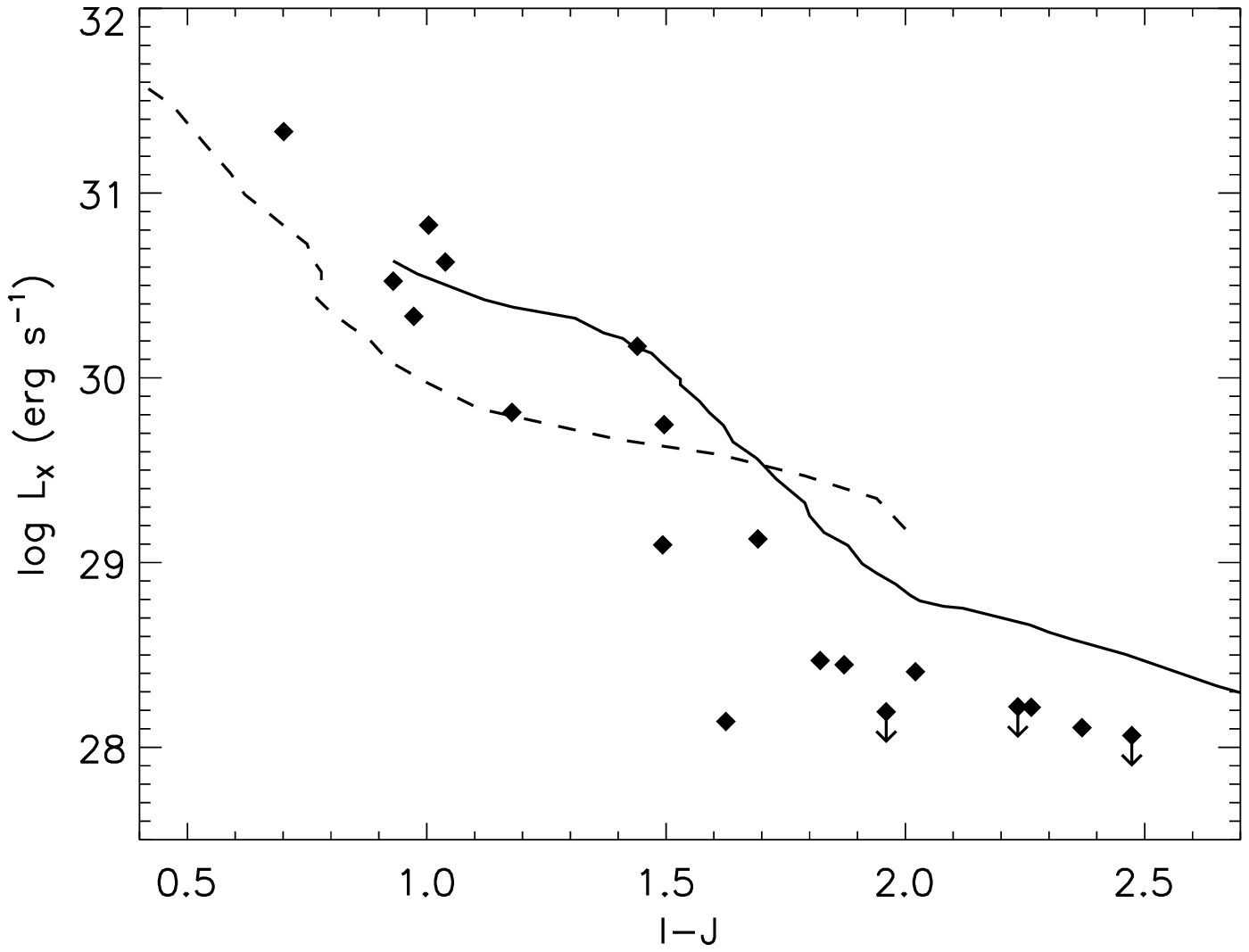}}
\caption{X-ray luminosity of USco members vs. $H-K$ ({\it left panel}) and $I-J$ ({\it right panel}) colors obtained from the 2MASS and DENIS catalogs. Dashed and solid lines represent the saturated level ($\log L_{\rm X}/L_{\rm bol}=-3$) predicted by the \protect{\citet{SiessDufour2000}} and \protect{\citet{BaraffeChabrier1998}} models, respectively.}
\label{fig:lxvscol}
\end{figure*}

\section{Discussion and conclusions}
\label{disc}

In the analyzed {\it XMM-Newton} observations we have detected and identified 22 photometric USco members with masses ranging from $\sim0.05$ to $\sim2\,M_{\sun}$. This stellar sample includes 1 CTTS and 7 WTTSs, while no indication is available for the remaining 14 members. Since the CTTSs frequency in the USco association is very low ($\sim2-5\%$) \citep{WalterVrba1994,Martin1998}, it is conceivable that source~132 is the only CTTS in our sample. In the following discussion the sample of 22 USco members detected in this X-ray survey is considered as a whole, since no significant differences between the X-ray properties of the CTTS source~132 and the other members have been found.

\subsection{X-ray luminosity}

We have explored how the X-ray luminosity of the USco members changes with NIR colors. We plot in Fig.~\ref{fig:lxvscol} $L_{\rm X}$ vs. NIR colors obtained from the 2MASS and DENIS counterparts. We also show in the two plots the locus of X-ray emission at the saturated level, $\log L_{\rm X}/L_{\rm bol}=-3$, predicted using the evolutionary models of \citet{SiessDufour2000} and \citet{BaraffeChabrier1998}.

\begin{figure}
\resizebox{\hsize}{!}{\includegraphics{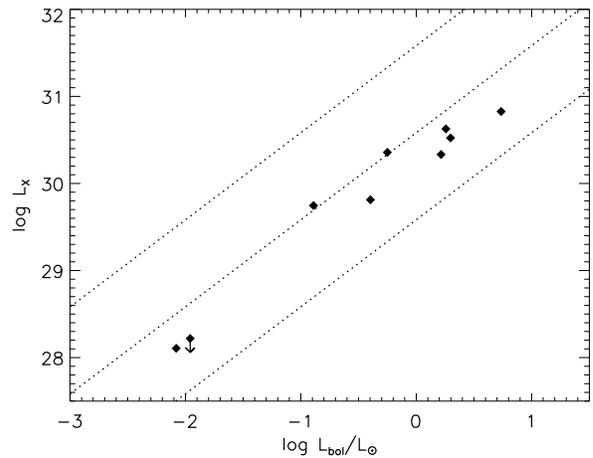}}
\caption{Fractional X-ray luminosity vs. bolometric luminosity. Dotted lines represent X-ray luminosity of $10^{-2}$, $10^{-3}$, and $10^{-4}$ with respect to bolometric luminosity.}
\label{fig:lxvslbol}
\end{figure}

For some of the X-ray detected USco members stellar parameters are available from literature: we list them in Table~\ref{tab:stelpar}. For this subset of USco members we plot the X-ray luminosity vs. bolometric luminosity (Fig.~\ref{fig:lxvslbol}), confirming the saturated X-ray emission of USco stars.

\begin{table}[b]
\begin{center}
\caption{Stellar parameters.}
\label{tab:stelpar}
\begin{tabular}{clrrc}
\hline\hline
\multicolumn{1}{c}{Source} & \multicolumn{1}{c}{Simbad} & \multicolumn{1}{c}{$\log L_{\rm bol}/L_{\sun}$} & Ref \\
\hline
  7                        & V* V1144 Sco               &  $-0.25$\hspace{0.5em}                          & W94 \\
 43                        & V* V1146 Sco               &  $-0.397$                                       & P99 \\
 45                        & DENIS-P J155601.0-233808   &  $-2.08$\hspace{0.5em}                          & B04 \\
 95                        & V* V1145 Sco               &  $-0.89$\hspace{0.5em}                          & W94 \\
114                        & GSC 06793-00569            &  $0.215$                                        & P99 \\
132                        & GSC 06793-00819            &  $0.736$                                        & P99 \\
158                        & GSC 06793-00994            &  $0.297$                                        & P99 \\
180                        & GSC 06793-00797            &  $0.257$                                        & P99 \\
                           & UScoCTIO 104               &  $-1.959$                                       & M04 \\

\hline
\end{tabular}
\end{center}
In the reference column W94, P99, B04 and M04 designate \citet{WalterVrba1994}, \citet{PreibischZinnecker1999}, \citet{Bouy2004} and \citet{MohantyJayawardhana2004}, respectively.
\end{table}

\subsection{Coronal plasma properties}

We evaluated the effective coronal temperature\footnote{Defined as weighted average of the best fit temperatures, adopting as weight the relevant emission measures.} of stars whose spectra were analyzed, which ranges from 6 to 40\,MK ($\log T ({\rm K}) = 6.8-7.6$). These values are compatible with that of the WTTSs of the younger $\rho$~Oph cloud \citep{OzawaGrosso2005}, of the L1551 region \citep{FavataGiardino2003}, and of the Orion Molecular Cloud 2 and 3 \citep{TsujimotoKoyama2002}. It suggests that the characteristic temperature of the plasma responsible for the X-ray emission in WTTS does not evolve significantly during the PMS stages. PMS stars in the Orion Nebula Cluster show increasing coronal temperature for increasing photospheric temperature \citet{PreibischKim2005}. Instead our results do not suggest any correlation of coronal temperatures with NIR colors, possibly because of the smaller stellar sample. 

All the USco stars, whose X-ray spectra were analyzed, showed elemental abundances reduced with respect to the solar photospheric values. The global metallicity is in all cases $\le0.35$ in solar units. Moreover Ne, when estimated independently from Fe, resulted less depleted than Fe. The Ne/Fe ratio ranges between 2 and 17 times the solar photospheric ratio.

As in the case of coronal temperatures, also coronal metallicity does not show any significant trend with respect to NIR colors. Hence the photospheric temperature does not seem to affect the metal depletion of coronal plasma.

\subsection{Short term X-ray variability}

We have found that 13, over the 22 photometric USco members, were variable during the observations. Since the 13 variable sources are also the  X-ray brightest, our result is likely biased by the quality of the X-ray data collected.

The analysis of the four strongest and isolated flares produced by USco members have displayed very hot plasmas with temperatures as high as 70\,MK, and peak luminosity up to $8\times10^{31}\,{\rm erg\,s^{-1}}$. We have derived the flaring loops semi-length finding that in three cases these sizes are slightly smaller than the stellar radius ($L\sim0.7-0.8\,R_{\star}$), while in the last case the flare originated in a compact loop with semi-length of $L\lesssim0.1\,R_{\star}$. These results are supported by studies on X-ray flares performed adopting different methods \citep{FavataMicela2001,ImanishiNakajima2003,GrossoMontmerle2004}. Hence coronal loops longer than the stellar radius, as those found in the Orion Nebula Cluster by \citet{FavataFlaccomio2005}, are rare and/or present only in the very early stages of stellar evolution. However the flares studied by \citet{FavataFlaccomio2005} have very long duration (with time decay of $\sim10-10^{2}$\,ks), and they may escape detection in observations with short exposure times.

The intermediate mass star HD~142578 (source 100, the most X-ray luminous in Fig.~\ref{fig:lxvscol}) displays an X-ray luminosity, evaluated over the entire observation, near the saturated level, which is not expected for an A2 young star. During the observation a large flare occurred, with the count rate rising up to a factor 100 in a fews ks. The X-ray luminosity evaluated in the quiescent phase is $\log L_{\rm X}=29.8$. Assuming that the X-ray emitter is an unresolved low-mass companion characterized by a saturated X-ray emission (hence $L_{\rm bol}=0.16\,L_{\odot}$), the \citet{SiessDufour2000} model predicts for it $M_{\star}\sim0.4\,M_{\odot}$ and $R_{\star}\sim1\,R_{\odot}$. Under this hypothesis the length of the flaring loop should have been $2\,R_{\star}$, and the X-ray luminosity during the flare should have reached $\sim10\%$ of the bolometric luminosity. Such high X-ray to bolometric ratios during flares have been already observed \citep[e.g.][]{FavataReale2000}.

The flare analysis of source 7 and 100 has shown that the plasma metallicity is higher in the flaring structures than in the quiet coronal plasma. Abundance variations during flares were observed in some cases \citep{TsuboiKoyama1998,GudelLinsky1999,FavataSchmitt1999,GudelAudard2001,ImanishiTsujimoto2002}, where metal enrichment up to a factor $\sim5$ was registered. The metallicity of the flaring plasma of source~100 resulted enhanced by a factor $\sim20$ with respect to the quiescent plasma.

The general view suggested for the abundance variation during flares invokes evaporation of photospheric material which then fills the flaring structures. If the mechanism, responsible for the FIP or inverse FIP effect in quiescent coronal structures, acts on time scales longer than the characteristic rise times of coronal flares (few ks) it is conceivable that the evaporated photospheric material radiates X-ray emission before being affected by any FIP-related effect. Therefore the abundances of the flaring structures may reflect the chemical composition of the underlying stellar photosphere.

Under this scenario we infer that the quiescent corona is actually metal depleted with respect to the underlying photosphere; on the other hand, since the flaring composition reaches the highest value of ${\rm Fe/Fe_{\odot}}\sim 0.5-1$, we conclude that PMS stars of the USco association do have solar-like composition.

Note that the abundance increase during strong flares is followed by an abundance decrease just after the flare maximum. This effect is visible in the flare analysis of sources~7 and 100 reported in Fig.~\ref{fig:fitres100}. In both cases the abundance decays on time scales of $\sim1-2$\,ks, much faster than the characteristic time (36\,ks) of the abundance decay observed in a flare of Algol \citep{FavataSchmitt1999}, which however showed much longer decay time (50\,ks). While it is conceivable that the rapid evaporation of photospheric material may produce the observed rise of metallicity on very short time scales, it is difficult to explain how the newly evaporated plasma is so efficiently depleted of heavy elements on so short time scales. 

\begin{acknowledgements}

We would like to thank F.~Reale for the useful discussion on flare analysis and I.~Pillitteri for helpful suggestions on the procedure of X-ray source detection. CA, EF, AM, GM, GP, and SS acknowledge for this work partial support from contract ASI-INAF I/023/05/0 and from the Ministero dell'Istruzione, Universit\`a e Ricerca. Based on observations obtained with {\it XMM-Newton}, an ESA science mission with instruments and contributions directly funded by ESA Member States and NASA.

\end{acknowledgements}

\bibliographystyle{aa}
\bibliography{upsco.bib}

\appendix

\section{Results of the flare analysis}
  
\begin{table}[h]
\renewcommand{\baselinestretch}{1.3}
\begin{center}
\caption{Time resolved analisys of the flare of source 7. Fits were performed with $N_{\rm H}=0.32\times10^{21}\,{\rm cm^{-2}}$. Unabsorbed X-ray luminosity is computed in the $0.5-8.0$\,keV band and takes into account the whole (flaring + quiescent) X-ray emission.}
\label{tab:flare7}
\footnotesize
\begin{tabular}{c@{\hspace{1mm}}c@{\hspace{1mm}}c@{\hspace{1mm}}c@{\hspace{1mm}}c}
\hline\hline
\multicolumn{1}{c}{Int.} & \multicolumn{1}{c}{$T ({\rm MK})$} & \multicolumn{1}{c}{${\it EM} ({\rm 10^{52} cm^{-3}})$} & \multicolumn{1}{c}{${\rm Fe/Fe_{\sun}}$} & $\log L_{\rm X} ({\rm erg\,s^{-1}})$ \\
\hline
   q1 &                                 $   8.5^{+   0.7}_{-   0.8}$ &                                 $  15.8^{+   3.0}_{-   3.3}$ &                                 $  0.10^{+  0.05}_{-  0.03}$ &   $ 29.89$ \\
   r1 &                                                  $\ge   8.6$ &                                 $  10.5^{+  11.2}_{-  10.5}$ &                                 $  0.17^{+ 24.54}_{-  0.17}$ &   $ 30.19$ \\
   d1 &                                 $  27.2^{+  12.4}_{-   8.4}$ &                                 $  37.6^{+  14.1}_{-  13.8}$ &                                 $  1.01^{+  2.30}_{-  0.68}$ &   $ 30.77$ \\
   d2 &                                 $  23.7^{+   9.5}_{-   5.0}$ &                                 $  32.0^{+   7.4}_{-   8.4}$ &                                 $  0.41^{+  0.70}_{-  0.27}$ &   $ 30.60$ \\
   d3 &                                 $  21.8^{+   6.0}_{-   5.4}$ &                                 $  29.1^{+   6.1}_{-   4.9}$ &                                 $  0.16^{+  0.22}_{-  0.14}$ &   $ 30.50$ \\
   q2 &                                 $  17.8^{+   1.7}_{-   1.7}$ &                                 $  30.4^{+   3.2}_{-   3.1}$ &                                 $  0.14^{+  0.08}_{-  0.06}$ &   $ 30.47$ \\
\hline
\end{tabular}
\normalsize
\end{center}
\end{table}

\begin{table}[h]
\renewcommand{\baselinestretch}{1.3}
\begin{center}
\caption{Time resolved analisys of the flare of source 100. Fits were performed with $N_{\rm H}=1.3\times10^{21}\,{\rm cm^{-2}}$. Unabsorbed X-ray luminosity is computed in the $0.5-8.0$\,keV band and takes into account the whole (flaring + quiescent) X-ray emission.}
\label{tab:flare100}
\footnotesize
\begin{tabular}{c@{\hspace{1mm}}c@{\hspace{1mm}}c@{\hspace{1mm}}c@{\hspace{1mm}}c}
\hline\hline
\multicolumn{1}{c}{Int.} & \multicolumn{1}{c}{$T ({\rm MK})$} & \multicolumn{1}{c}{${\it EM} ({\rm 10^{52} cm^{-3}})$} & \multicolumn{1}{c}{${\rm Fe/Fe_{\sun}}$} & $\log L_{\rm X} ({\rm erg\,s^{-1}})$ \\
\hline
   q1 &                                 $   8.6^{+   1.0}_{-   0.9}$ &                                 $  18.8^{+   3.7}_{-   3.4}$ &                                 $  0.02^{+  0.02}_{-  0.01}$ &   $ 29.79$ \\
   r1 &                                 $  62.4^{+  27.1}_{-  12.6}$ &                                 $  77.7^{+  10.2}_{-   9.8}$ &                                 $  0.42^{+  0.56}_{-  0.42}$ &   $ 31.08$ \\
   r2 &                                 $  68.6^{+  12.7}_{-  10.6}$ &                                 $ 338.0^{+  27.0}_{-  27.0}$ &                                 $  0.43^{+  0.36}_{-  0.31}$ &   $ 31.71$ \\
   r3 &                                 $  51.9^{+   6.2}_{-   6.1}$ &                                 $ 516.4^{+  33.8}_{-  33.6}$ &                                 $  0.47^{+  0.24}_{-  0.22}$ &   $ 31.86$ \\
   p1 &                                 $  38.8^{+   4.7}_{-   2.6}$ &                                 $ 654.9^{+  36.3}_{-  40.5}$ &                                 $  0.32^{+  0.16}_{-  0.12}$ &   $ 31.89$ \\
   p2 &                                 $  34.6^{+   2.5}_{-   2.5}$ &                                 $ 708.6^{+  36.6}_{-  36.3}$ &                                 $  0.14^{+  0.10}_{-  0.09}$ &   $ 31.87$ \\
   d1 &                                 $  33.1^{+   2.1}_{-   2.7}$ &                                 $ 686.2^{+  28.0}_{-  27.6}$ &                                 $  0.17^{+  0.08}_{-  0.07}$ &   $ 31.85$ \\
\hline
\end{tabular}
\normalsize
\end{center}
\end{table}

\begin{table}[h]
\renewcommand{\baselinestretch}{1.3}
\begin{center}
\caption{Time resolved analisys of the flare of source 132. Fits were performed with $N_{\rm H}=1.6\times10^{21}\,{\rm cm^{-2}}$. Unabsorbed X-ray luminosity is computed in the $0.5-8.0$\,keV band and takes into account the whole (flaring + quiescent) X-ray emission.}
\label{tab:flare132}
\footnotesize
\begin{tabular}{c@{\hspace{1mm}}c@{\hspace{1mm}}c@{\hspace{1mm}}c@{\hspace{1mm}}c}
\hline\hline
\multicolumn{1}{c}{Int.} & \multicolumn{1}{c}{$T ({\rm MK})$} & \multicolumn{1}{c}{${\it EM} ({\rm 10^{52} cm^{-3}})$} & \multicolumn{1}{c}{${\rm Fe/Fe_{\sun}}$} & $\log L_{\rm X} ({\rm erg\,s^{-1}})$ \\
\hline
   p1 &                                 $  34.3^{+   6.2}_{-   4.8}$ &                                 $  29.5^{+   2.7}_{-   2.7}$ &                                                    $=  0.21$ &   $ 30.92$ \\
   d1 &                                 $  29.2^{+   8.1}_{-   5.0}$ &                                 $  22.7^{+   2.7}_{-   2.9}$ &                                                    $=  0.21$ &   $ 30.87$ \\
   d2 &                                 $   8.7^{+   2.7}_{-   0.7}$ &                                 $  14.1^{+   2.4}_{-   2.4}$ &                                                    $=  0.21$ &   $ 30.80$ \\
   d3 &                                 $   8.0^{+   1.5}_{-   0.6}$ &                                 $   7.7^{+   1.9}_{-   1.9}$ &                                                    $=  0.21$ &   $ 30.76$ \\
   d4 &                                 $   7.8^{+   1.7}_{-   4.4}$ &                                 $   3.8^{+   2.1}_{-   1.9}$ &                                                    $=  0.21$ &   $ 30.74$ \\
   r1 &                                 $   7.9^{+   1.9}_{-   3.6}$ &                                 $   3.6^{+   2.3}_{-   3.6}$ &                                                    $=  0.21$ &   $ 30.74$ \\
   d5 &                                 $   7.9^{+   0.9}_{-   2.1}$ &                                 $   4.7^{+   2.0}_{-   1.8}$ &                                                    $=  0.21$ &   $ 30.75$ \\
\hline
\end{tabular}
\normalsize
\end{center}
\end{table}

\begin{table}[h]
\renewcommand{\baselinestretch}{1.3}
\begin{center}
\caption{Time resolved analisys of the flare of source 173. Fits were performed with $N_{\rm H}=1.2\times10^{21}\,{\rm cm^{-2}}$. Unabsorbed X-ray luminosity is computed in the $0.5-8.0$\,keV band and takes into account the whole (flaring + quiescent) X-ray emission.}
\label{tab:flare173}
\footnotesize
\begin{tabular}{c@{\hspace{1mm}}c@{\hspace{1mm}}c@{\hspace{1mm}}c@{\hspace{1mm}}c}
\hline\hline
\multicolumn{1}{c}{Int.} & \multicolumn{1}{c}{$T ({\rm MK})$} & \multicolumn{1}{c}{${\it EM} ({\rm 10^{52} cm^{-3}})$} & \multicolumn{1}{c}{${\rm Fe/Fe_{\sun}}$} & $\log L_{\rm X} ({\rm erg\,s^{-1}})$ \\
\hline
   p1 &                                 $  22.5^{+   3.2}_{-   4.0}$ &                                 $  75.5^{+   5.4}_{-   5.4}$ &                                                    $=  0.35$ &   $ 30.90$ \\
   d1 &                                 $  21.6^{+   2.3}_{-   2.2}$ &                                 $  47.5^{+   2.9}_{-   2.9}$ &                                                    $=  0.35$ &   $ 30.72$ \\
   d2 &                                 $  18.0^{+   1.5}_{-   1.2}$ &                                 $  28.6^{+   2.0}_{-   1.9}$ &                                                    $=  0.35$ &   $ 30.54$ \\
   d3 &                                 $  14.2^{+   0.8}_{-   1.3}$ &                                 $   7.4^{+   0.6}_{-   0.8}$ &                                                    $=  0.35$ &   $ 30.19$ \\
   q1 &     $   8.7^{+   0.6}_{-   1.2},   18.6^{+  16.8}_{-  11.2}$ &     $   5.0^{+   0.8}_{-   1.4},    3.9^{+   1.0}_{-   1.0}$ &                                                    $=  0.35$ &   $ 29.94$ \\
   r1 &                                 $ 159.7^{+ 400.8}_{- 107.7}$ &                                 $   3.0^{+   0.5}_{-   0.6}$ &                                                    $=  0.35$ &   $ 30.13$ \\
   d4 &                                 $  27.4^{+  51.4}_{-  12.0}$ &                                 $   3.3^{+   1.1}_{-   1.1}$ &                                                    $=  0.35$ &   $ 30.08$ \\
\hline
\end{tabular}
\normalsize
\end{center}
\end{table}

\Online

\section{CTIO source list}

A preprint version of the paper containing the CTIO source list is available at {\tt  www.astropa.unipa.it/Library/preprint.html}

\section{Danish 1.54m source list}

A preprint version of the paper containing the Danish 1.54m source list is available at {\tt www.astropa.unipa.it/Library/preprint.html}

\section{XMM-Newton source list}
\scriptsize
\begin{longtable}{cccr@{ $\pm$ }lrrcccc}
\caption{\label{tab:src} X-ray source detected in the two observations of Upper Sco.} \\
\hline\hline
 \multicolumn{1}{c}{Source} & \multicolumn{1}{c}{RA}      & \multicolumn{1}{c}{DEC}     & \multicolumn{2}{c}{PN Count rate$^{\rm a}$} & \multicolumn{1}{c}{$XC1^{\rm b}$} & \multicolumn{1}{c}{$XC2^{\rm b}$} & \multicolumn{2}{c}{2MASS}                                         & \multicolumn{2}{c}{DENIS}                                         \\
                            & \multicolumn{1}{c}{(J2000)} & \multicolumn{1}{c}{(J2000)} & \multicolumn{2}{c}{${\rm (cts\,ks^{-1})}$}  &                                   &                                   & \multicolumn{1}{c}{Source} & \multicolumn{1}{c}{Offset (\arcsec)} & \multicolumn{1}{c}{Source} & \multicolumn{1}{c}{Offset (\arcsec)} \\
\hline
\endfirsthead
\caption{continued.} \\
\hline\hline
 \multicolumn{1}{c}{Source} & \multicolumn{1}{c}{RA}      & \multicolumn{1}{c}{DEC}     & \multicolumn{2}{c}{PN Count rate$^{\rm a}$} & \multicolumn{1}{c}{$XC1^{\rm b}$} & \multicolumn{1}{c}{$XC2^{\rm b}$} & \multicolumn{2}{c}{2MASS}                                         & \multicolumn{2}{c}{DENIS}                                         \\
                            & \multicolumn{1}{c}{(J2000)} & \multicolumn{1}{c}{(J2000)} & \multicolumn{2}{c}{${\rm (cts\,ks^{-1})}$}  &                                   &                                   & \multicolumn{1}{c}{Source} & \multicolumn{1}{c}{Offset (\arcsec)} & \multicolumn{1}{c}{Source} & \multicolumn{1}{c}{Offset (\arcsec)} \\
\hline
\endhead
\hline
\endfoot
\hline
\multicolumn{11}{p{0.89\textwidth}}{

\normalsize
$^{\rm a}$~PN count rate that would be observed if sources were located on the telescope axis. 
 
$^{\rm b}$~X-ray colors, derived from PN, defined as $XC1=2.5\log(M/S)$ and $XC2=2.5\log(H/M)$ where the energy range of PN has been divided into the bands: soft $0.3-0.85$, medium $0.85-1.4$, and hard $1.4-7.9$\,keV. $S$, $M$ and $H$ indicate the number of photons in the relevant energy band. 
 
$^{\rm c}$~X-ray colors derived from MOS instruments for those sources which fall outside the PN field of view. 
\scriptsize
}
\endlastfoot
    1 &           15:56:12.4 &          -23:51:04.5 &      22.4 &       2.4 &                      $   0.5^{\rm c}             $ &                      $  -3.2^{\rm c}             $ &     15561238-2351030 &   1.5 &     J155612.3-235102 &   1.6 \\
    1 &                      &                      & \multicolumn{2}{c}{ } &                                                    &                                                    &     15561210-2351065 &   4.0 &                      &       \\
    2 &           15:56:17.2 &          -23:50:56.5 &      22.3 &       2.4 &                      $   1.8^{\rm c}             $ &                      $   0.7^{\rm c}             $ &     15561698-2350589 &   3.5 &     J155616.9-235058 &   3.6 \\
    3 &           15:56:20.0 &          -23:49:46.9 &       6.1 &       0.9 &                      $   0.5^{\rm\hspace{0.32em}}$ &                      $   0.3^{\rm\hspace{0.32em}}$ &                      &       &                      &       \\
    4 &           15:57:00.6 &          -23:49:42.7 &       2.3 &       0.8 &             $\cdot\cdot\cdot^{\rm\hspace{0.32em}}$ &                      $   1.5^{\rm\hspace{0.32em}}$ &                      &       &                      &       \\
    5 &           15:56:43.8 &          -23:49:25.9 &      11.5 &       1.3 &                      $  -0.6^{\rm\hspace{0.32em}}$ &                      $   0.6^{\rm\hspace{0.32em}}$ &                      &       &                      &       \\
    6 &           15:55:52.5 &          -23:48:45.3 &      17.6 &       2.0 &                      $   0.6^{\rm c}             $ &                      $   0.1^{\rm c}             $ &     15555254-2348460 &   0.8 &     J155552.5-234846 &   0.9 \\
    7 &           15:56:29.4 &          -23:48:20.5 &     439.4 &       6.1 &                      $  -0.2^{\rm\hspace{0.32em}}$ &                      $  -0.7^{\rm\hspace{0.32em}}$ &     15562941-2348197 &   0.8 &                      &       \\
\end{longtable}
\normalsize

\end{document}